\documentclass[pra, twocolumn, notitlepage, superscriptaddress,10pt]{revtex4-1}
\usepackage{gpkmac}

\begin{document}

\title{Quantum Zeno Effect and the Many-body Entanglement Transition}
\author{Yaodong Li}
\affiliation{Department of Physics, University of California, Santa Barbara, CA 93106, USA}
\author{Xiao Chen}
\affiliation{Kavli Institute for Theoretical Physics, University of California, Santa Barbara, CA 93106, USA}
\author{Matthew P. A. Fisher}
\affiliation{Department of Physics, University of California, Santa Barbara, CA 93106, USA}

\date{November 2, 2018}

\begin{abstract}

    We introduce and explore a one-dimensional ``hybrid" quantum circuit model consisting
  of both unitary gates and projective measurements.  While the unitary gates are drawn from a
random distribution and act uniformly in the circuit, the measurements
  are made at random positions and times throughout the system.
   By varying the measurement rate we
  can tune between the volume law entangled phase for
  the random unitary circuit
  model (no measurements) and a ``quantum Zeno phase" where strong measurements suppress the entanglement growth to saturate in an area-law.
  Extensive numerical simulations of the quantum trajectories of the many-particle wavefunctions (exploiting Clifford circuitry to access systems
  up to 512 qubits) provide evidence for a stable ``weak measurement phase" that exhibits
  volume-law entanglement entropy, with a coefficient decreasing with increasing measurement rate.  We also present evidence for a novel continuous quantum dynamical phase transition between the ``weak measurement phase" and the ``quantum Zeno phase", driven
  by a competition between the entangling tendencies of unitary evolution and the disentangling tendencies of projective measurements.
  Detailed steady-state and dynamic critical properties of this novel quantum entanglement transition are accessed.

\end{abstract}

\maketitle


\section{Introduction}

Entanglement provides a convenient characterization for both stationary states and unitary dynamics of quantum systems, especially those with no symmetry. For many-body systems, the entanglement entropy
is useful in
classifying both ground and excited states~\cite{Kitaev2006, Levin2006, Calabrese2004, Ryu2006, Casini2012, Page1993, Goldstein2006}.
Entanglement dynamics under unitary time evolution in a driven system or after a quench
also exhibits universal behavior.  From an unentangled state, the entanglement entropy grows linearly in time, saturating with a volume-law.
For integrable systems the emission of entangled pairs of quasiparticles~\cite{Calabrese_2005,Calabrese_2007}
offers a convenient picture, but chaotic systems also exhibit similar entanglement growth~\cite{Kim2013,Mezei2017},
as illustrated in random unitary circuit models~\cite{nahum2017KPZ, nahum2018operator, keyserlingk2018operator}.

Quantum systems subjected to both (continuous) measurements and unitary dynamics offer another class of quantum dynamical behavior, described in terms of ``quantum trajectories",
and well explored in the context of few qubit systems~\cite{Wiseman1996},
{
quantum spin systems~\cite{dasgupta2016measurement}, and trapped ultracold atoms~\cite{elliott2015cold, elliott2016cold}
}.
With strong and continuous
measurements the state vector can become localized in the Hilbert space,
an example of a quantum Zeno effect~\cite{Misra1977zeno}.
While the related quantum dynamics has been explored in many-body ``open" systems~\cite{Breuer2002theory}, which focus on the mixed state density matrix and can be described by Lindblad equations, this formalism does not offer access to quantum entanglement.
Very recently in Ref.~\cite{cao2018monitoring}, the entanglement dynamics of quantum state trajectories
have been explored
for (an integrable) model of non-interacting fermions subjected to continuous measurements
of local occupancy, performed at a constant rate throughout the system.  Remarkably, authors of Ref.~\cite{cao2018monitoring} conclude that the late time entanglement entropy growth
saturates to an area law for arbitrarily weak measurement.   A ``weak measurement phase"
with volume law entanglement is not present.

In this work we explore the many-body dynamics of a ``hybrid" one-dimensional quantum circuit model consisting
of both (projective) measurements and unitary gates, depicted in Fig.~\ref{fig1}.
The dynamics of quantum entanglement can be accessed by following quantum trajectories of the many-body wavefunctions.  We primarily focus on
circuits that consist of measurements that are made at random positions and times throughout the system and have unitary gates, chosen from a random distribution, that act uniformly in the circuit.  While being a discrete-time generalization
of the model in Ref.~\cite{cao2018monitoring}, this model is non-integrable.
By varying the measurement rate we
can tune between the volume law entangled phase for
the random unitary circuit
model (no measurements)~\cite{nahum2018operator, keyserlingk2018operator} and a strong measurement ``quantum Zeno regime''
where entanglement growth is suppressed, saturating in an area-law.
We analyze the quantum trajectories numerically both for random Haar and random Clifford
unitaries, which are minimal models describing chaotic non-integrable systems.  In the latter case, by restricting measurements to the Pauli group,
we can access the long-time quantum dynamics of very large systems, up to 512 qubits.
Our numerics supports several striking conclusions.

Firstly, we find that volume law entanglement of the random unitary model survives ``weak" measurements,
but with the coefficient of the volume law decreasing with increasing measurement strength - a {stable} ``weak measurement phase".
The absence of this phase in the integrable model of Ref.~\cite{cao2018monitoring} is
perhaps due to the effectiveness of
(local) measurements in removing long-ranged entanglement
from their highly susceptible EPR pairs.
In contrast, the entanglement in chaotic systems is encoded in the sign structure
of the (essentially) random wavefunction~\cite{Grover2015}, evidently less disturbed by measurement.

Secondly, we present numerical evidence that upon increasing the rate of the measurement $p$,
this ``weak measurement phase"
with volume-law entanglement, undergoes a continuous dynamical quantum phase
transition into an area-law entangled ``quantum Zeno phase".  While we do not have analytic access to this transition, our
data can be collapsed into a standard finite-size scaling form: the entanglement entropy,
$S_A(p, L_A)$,  with subsystem size $L_A$ and $p-p_c$ the deviation of the measurement rate
from criticality,
fits a form $S_A(p, L_A) = L_A^\gamma F \( (p-p_c) L_A^{1/\nu} \)$, with scaling function $F$.
Right at criticality, $p=p_c$, the entanglement grows with system size as a sub-linear power law, $S_A(p_c, L_A) \sim L_A^\gamma$, with $\gamma \approx 1/3$ - intermediate between volume and area law scaling.  Moreover, we investigate the entanglement entropy dynamics starting from an initial product state with no entanglement. We find that when $p=p_c$ the entanglement grows as a sub-linear power law in time.

The rest of this article is organized as follows.
In Section~\ref{sec2}, we define the circuit model.
In Section~\ref{sec3}, we discuss dynamics of the entanglement entropy and provide numerical evidence for the entanglement transition.
We conclude with discussions in Section~\ref{sec5}.


\section{The circuit model \label{sec2}}
\begin{figure}[t]
    \centering
    \includegraphics[width=.5\textwidth]{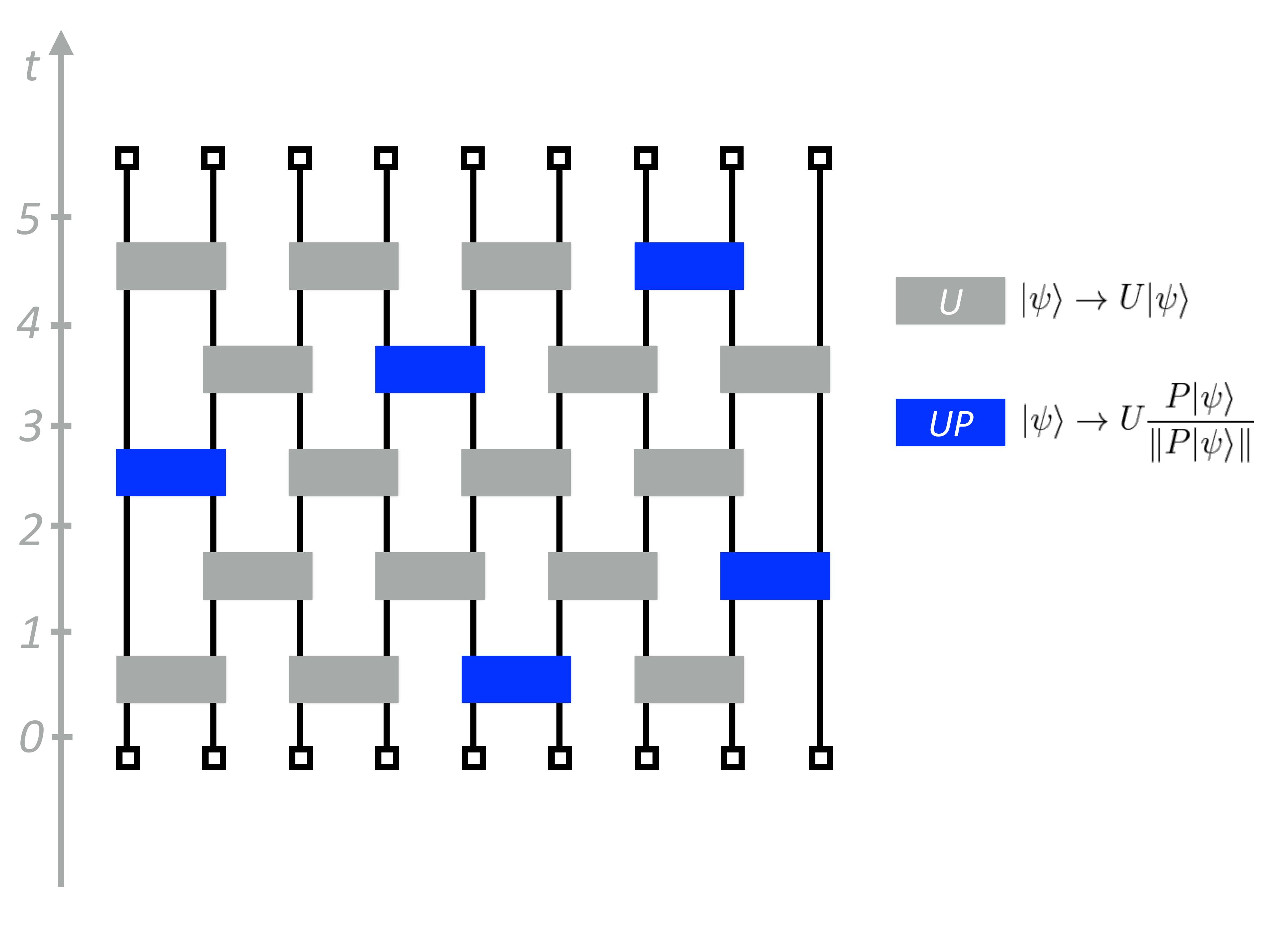}
    \caption{The structure of the hybrid circuit model. In this paper we will focus on 1D circuits with nearest neighbor gates. Each site has a spin-1/2 degree of freedom, and each block represents a gate operation on two qubits. See main text for details.}
    \label{fig1}
\end{figure}

We consider a setup with one-dimensional geometry, where the qubits are arranged on a chain of $L$ sites, with one qubit on each site.
The dynamics of the system is governed by the quantum circuit with ``brick-layer'' structure, see
Fig.~\ref{fig1}. The circuit is composed of quantum gates on pairs of neighboring qubits, whose pattern
of arrangement is periodic in the time direction. Each discrete time period of the circuit contains two
layers, and each layer has $L/2$ gates, acting on all the {odd} links in the first layer, and all the {even} links in the second.
Throughout the paper we will assume open boundary conditions on the circuit.

In Fig.~\ref{fig1}, we take each block to represent a gate operation. There are two different types of gates, as labelled by different colors.
Each gray block represents one unitary gate (denoted $U$), and acts on the states as
\begin{eqnarray}
	\label{eq1-1}
	\ket{\psi} \to U\ket{\psi},
\end{eqnarray}
while each blue block represents a ``unitary-projective'' gate, which performs a projective measurement before the unitary (which we denote $UP$), and acts on the state in the following fashion,
\begin{eqnarray}
    \label{eq1-2}
    \ket{\psi} \to U 
    \frac{P_\alpha \ket{\psi}}{\lVert P_\alpha \ket{\psi} \rVert},
\end{eqnarray}
where $\{P_\alpha\}$ is a complete, mutually exclusive set of projectors, for which $P_\alpha P_\beta = \delta_{\alpha\beta}P_\alpha$ and $\sum_\alpha P_\alpha = 1$. The outcome $\alpha$
happens at a probability given by Born's rule, $p_\alpha = \bra{\psi} P_\alpha \ket{\psi}$.
We need not specify the ordering of the gates within each layer since they commute with one another.
Notice that the unitary transformation $U$ can be chosen to be conditioned on the outcome of the measurement $\alpha$, which is approriate for the binding of symmetric {Posner} molecules~\cite{mpaf1707qds}, and plays a prominent role in the quantum brain scenario~\cite{mpaf1508qcog, swift2018posner, YaodongMPAF2018Cat}.
In that context, a similar circuit for quantum information processing in the Posner model appeared in \cite{halpern1711qiqcog}, {where  unitaries conditioned on the measurement outcomes are utilized in preparing resource states for universal measurement-based quantum computation.}

In Fig.~\ref{fig1}, we draw the circuit in such a way that the occurrences of the unitary-projective gates in space and time are random. For concreteness, we focus on the simple situation in which we independently choose each gate to be a $UP$ gate with probability $p$ and a $U$ gate with probability $1-p$. In the limit $p \to 0$, no projections are made in the circuit, and it reduces to a unitary circuit. In the limit $p \to 1$, a projective measurement is made before all of the unitaries.

When simulating these circuit models, we will primarily follow the evolution of a pure state wavefunction, as in Eq.~\eqref{eq1-1} and \eqref{eq1-2} - that is, a  {quantum trajectory}.
Alternatively, these quantum circuit models can be studied in terms of {quantum
channels}~\cite{nielsen2010qiqc}, by keeping track of the mixed state due to projections, rather than pure states given
by different instances of the measurement outcomes individually.
Generally, a quantum channel is a trace-preserving, completely positive map that
takes one density matrix to another. It has the operator-sum representation,
\begin{eqnarray}
    \rho \to \mc{E}[\rho] = \sum_{\alpha = 0}^{m-1} M^\pg_\alpha \rho M_\alpha^\dg,
\end{eqnarray}
where $\{M_\alpha\}$ are {Kraus operators} which satisfy the condition,
\begin{eqnarray}
    \sum_{\alpha=0}^{m-1} M_\alpha^\dg M^\pg_\alpha = 1.
\end{eqnarray}
In our case, we will take $\rho$ to be the density matrix of the system involving all measurement
outcomes, and the Kraus operators to be $M_\alpha = U P_\alpha$.

Using the notation introduced above, the time evolution of the density matrix as governed by the quantum
circuit in Fig.~\ref{fig1} is given by,
\begin{eqnarray}
    \rho(t) = \( \prod_{t^\p = 0}^{t-1} \mc{E}(t^\p) \) \lz \rho(0) \rz,
\end{eqnarray}
where $\rho(t)$ is the density matrix of the $L$-qubit system at time $t$, and
\begin{eqnarray}
    \mc{E}(t) = \begin{cases}
        \prod_{i \text{ odd }} \mc{E}_{i, i+1} (t), \text{ if $t$ is odd}, \\
        \prod_{i \text{ even}} \mc{E}_{i, i+1} (t), \text{ if $t$ is even}.
    \end{cases}
\end{eqnarray}
Each $\mc{E}_{i, i+1}(t)$ stands for a quantum channel acting locally on the pair of neighboring qubits $
(i, i+1)$ at time $t$.

We are interested in computing the discrete time quantum dynamics of these ``hybrid" unitary/projective circuits. Specifically, to characterize this dynamics, we will compute both
{thermal} and {entanglement} entropies.
The thermal entropy is extracted from the (evolving mixed state) density matrix,
\begin{eqnarray}
    \label{eq:Sth}
    S_{\rm th}(t) = \overline{s_2 \( \rho(t) \)},
\end{eqnarray}
where the overline denotes an average over all realizations of the random unitaries.
The entanglement entropy is defined for the pure state quantum trajectories, $\ket{\psi(t)}$, using the usual bipartition $(A, B)$ of the system, where $A$ is the subsystem of qubits $\{ 1, 2, \ldots, L_A\}$,
\begin{eqnarray}
    \label{eq:See}
    S_A(L_A, t) = \overline{\overline{s_2 \( \rho_A(t) \)}}, \quad \rho_A(t) = \textrm{Tr}_B \ket{\psi(t)} \bra{\psi(t)}.
\end{eqnarray}
The double overline denotes an average over all realizations of the
random unitaries, and over all possible positions and outcomes of the projective measurements.
Here, $s_2$ denotes the second Renyi entropy, defined for any density matrix (including $\rho$ and the reduced density matrix $\rho_A$),
\begin{eqnarray}
    s_2(\rho) \equiv -\log_2 \textrm{Tr} \rho^2.
\end{eqnarray}
Since the entanglement entropy will depend (very weakly) on the particular quantum trajectory
that follows from a set of measurement results, we will perform an average over many different ``runs'' through the circuit, averaging $s_2(\rho_A)$ over the ensemble of trajectories.

\subsection{Model Specifications \label{sec2-A}}

\begin{table*}[t]
\caption{Summary of models considered in this paper.}
\begin{center}
    \begin{tabular}{ | l | l | l | l | l | }
    \hline
    Model & Projectors & Unitaries & Rate of measurement $p$ & Section \\ \hline
    A1 & $\{P^{(1)}_{ab}\}, \ a, b\in\{0, 1\}$  & $U = \text{Haar}$ & $p=1$ & \ref{sec3-A}\\ \hline
    A2 & $\{P^{(2)}_{a}\}, \ a \in\{0, 1\}$     & $U = \text{Haar}$ & $p=1$ & \ref{sec3-A}\\ \hline
    B1 & $\{P^{(1)}_{ab}\},  \ a, b\in\{0, 1\}$ & $U = \text{Clifford}$  & $0 \le p \le 1$ & \ref{sec3-B}\\ \hline
    B2 & $\{P^{(2)}_{a}\}, \ a \in\{0, 1\}$     & $U = \text{Clifford}$& $0 \le p \le 1$ & \ref{sec3-B}\\ \hline
    \end{tabular}
\end{center}
\label{tab1}
\end{table*}

The behavior of the circuit model is entirely determined by the $U$, $P$ gates and the rate of measurement $p$.
For concreteness, in this paper we focus on two sets of $P$ gates on neighboring pairs of qubits.
\begin{enumerate}
\item
The first is a set of four rank-1 measurements,
\begin{eqnarray}
	\label{eq:P1_00}
	P^{(1)}_{00} &=& \ket{\ua\ua} \bra{\ua\ua}, \\
	\label{eq:P1_01}
	P^{(1)}_{01} &=& \ket{\ua \da} \bra{\ua\da}, \\
	\label{eq:P1_10}
	P^{(1)}_{10} &=& \ket{\da \ua} \bra{\da\ua}, \\
	\label{eq:P1_11}
	P^{(1)}_{11} &=& \ket{\da \da} \bra{\da\da}.
\end{eqnarray}
These can be effected by measuring both $Z_1$ and $Z_2$ operators on the two qubits, and have the convenient representation $P_{ab} = (1+(-1)^a Z_1)(1+(-1)^b Z_2)/4$.
After one such measurement, the two qubits are in a product state, and are completely disentangled from the other qubits.

Circuits with this set of rank-1 projectors can be considered as minimal discrete-time generalizations of the continuous monitoring model studied in Ref.~\cite{cao2018monitoring}, in which rank-1 onsite measurements of a 1D fermion chain are performed continuously in time at a constant rate.
However, there are also important differences between these two models.
While the quantum trajectory of the fermion model is described by a stochastic Schr\"{o}dinger equation, in the circuit model the quantum trajectory is discontinuous and has no straightforward description in terms of differential equations.
Moreover, the model in Ref.~\cite{cao2018monitoring} is integrable, and the entanglement entropy presumably comes from spatially separated EPR pairs, which are very susceptible to local measurements.
Indeed, the entanglement was found to saturate to an area law for any non-zero measurement rate.
In contrast, the time evolution of the circuit models is in general non-integrable, and does not have an obvious quasiparticle description. Thus, the entanglement dynamics can be quite different from that of an integrable model.

\item
Another set consists of two rank-2 measurements, taken to be a sum of the projectors in the previous set,
\begin{eqnarray}
	\label{eq:P2_0}
	P^{(2)}_0 &=& \ket{\ua \ua} \bra{\ua\ua} + \ket{\da\da} \bra{\da\da}, \\
	\label{eq:P2_1}
	P^{(2)}_1 &=& \ket{\ua \da} \bra{\ua\da} + \ket{\da\ua} \bra{\da\ua}.
\end{eqnarray}
These projectors represent the measurement of the operator $Z_1 Z_2$, and can be written as $P_a = (1 + (-1)^a Z_1 Z_2)/2$.
Since this set of measurements has rank-2, leaving some space for entanglement, they are expected to be less effective in suppressing the entanglement entropy.
Indeed, if one starts with a random Page state~\cite{Page1993}, after a row of $L/2$ $\{P^{(2)}\}$ projections in the $p=1$ limit, the state still has volume law entanglement, since it is a random Page state in $2^{L/2}$ dimensions.
For this reason, the interplay between $\{P^{(2)}\}$ and unitary time evolution is more delicate.
\end{enumerate}

For the local unitary gates that may or may not follow a projective measurement, we will consider the following two cases:
\begin{enumerate}[A.]
\item
All the unitaries, regardless of whether there is a preceding measurement and of the measurement outcome, are taken {independently} of each other from the Haar measure on $\mathsf{U}(4)$, that is the circular unitary ensemble (CUE)~\cite{mehta2004matrices, loggasrandommatrices}.
Since random Haar unitaries are quite generic, and can represent a large class of many-body chaotic models with local interaction~\cite{nahum2018operator, keyserlingk2018operator}, we expect our result for this model to be generic.

\item
All the unitaries, regardless of whether there is a preceding measurement and of the measurement outcome, are taken {independently} from the uniform distribution over the $2$-qubit Clifford group~\cite{gottesman9807heisenberg}, the latter being a finite group generated by the following unitaries,
\begin{eqnarray}
	\text{CNOT}_L &=& \frac{1}{2}(1+Z_1) + (1-Z_1)X_2,\\
	\text{CNOT}_R &=& \frac{1}{2}(1+Z_2) + (1-Z_2)X_1,\\
	H &=& \frac{1}{\sqrt{2}} (X+Z) = \frac{1}{\sqrt{2}} \begin{pmatrix}
		1 & 1 \\
		1 & -1
	\end{pmatrix}\\
	P &=& \sqrt{Z} = \begin{pmatrix}
		1 & 0\\
		0 & i
	\end{pmatrix}.
\end{eqnarray}

This model is very similar to the random Haar circuit, except that the Clifford unitaries can be simulated efficiently and thus scaled up to large system sizes~\cite{gottesman9807heisenberg, aaronson0406chp}.
In fact, random Clifford unitaries approximate the Haar random unitaries quite well, being known as a unitary 2-design~\cite{DiVincenzo2002hiding}.
Moreover, in the absence of any projective measurements, both the random Haar and random Clifford unitary circuits generate (maximal) volume law entanglement entropy~\cite{nahum2018operator}.
\end{enumerate}

In Table~\ref{tab1}, we summarize the four models that we will study, including their corresponding sections.

\section{Quantum Dynamics of Hybrid Circuits \label{sec3}}

In this section, we explore the quantum dynamical behavior of the entanglement entropy for the four circuit models defined in the table in Section~\ref{sec2}.
While the entanglement and operator dynamics in unitary Haar random circuits with $p=0$ has an emergent hydrodynamic description~\cite{nahum2018operator, keyserlingk2018operator},
we have not identified an analytic approach for nonzero $p$ and will resort to numerics.

 \subsection{The Quantum Zeno Phase \label{sec3-A}}

 Here we focus on the strong measurement limit $p=1$, that is models A1 and A2, in order to identify a ``Zeno phase'' in which the entanglement entropy saturates to an area law. Despite their differences, we expect (and find) that models A1 and A2 have qualitatively similar behaviors, which are presumably generic to other chaotic systems with strong measurements.

 \begin{figure}[t]
     \centering
     \includegraphics[width=.23\textwidth]{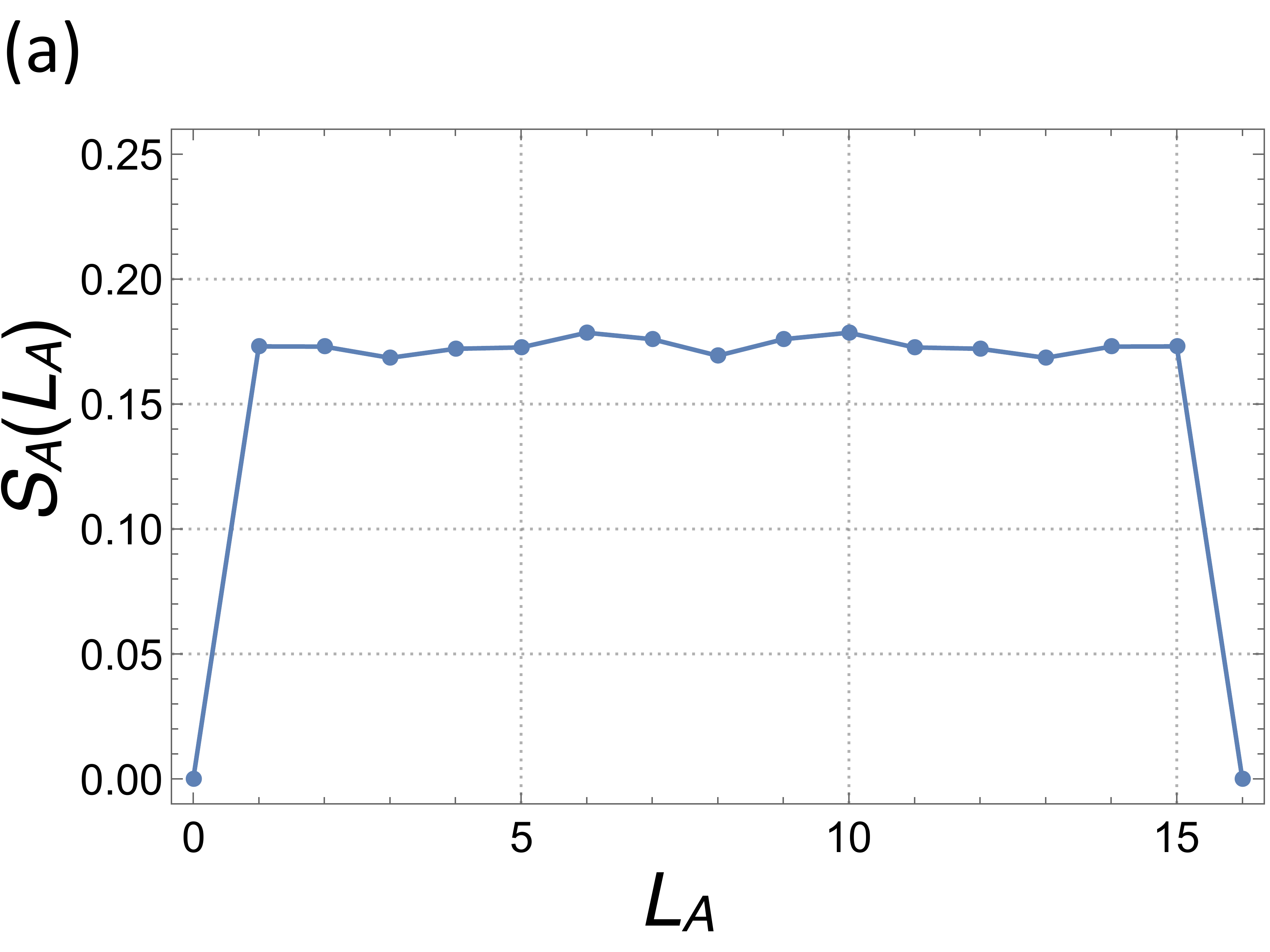}
     \includegraphics[width=.23\textwidth]{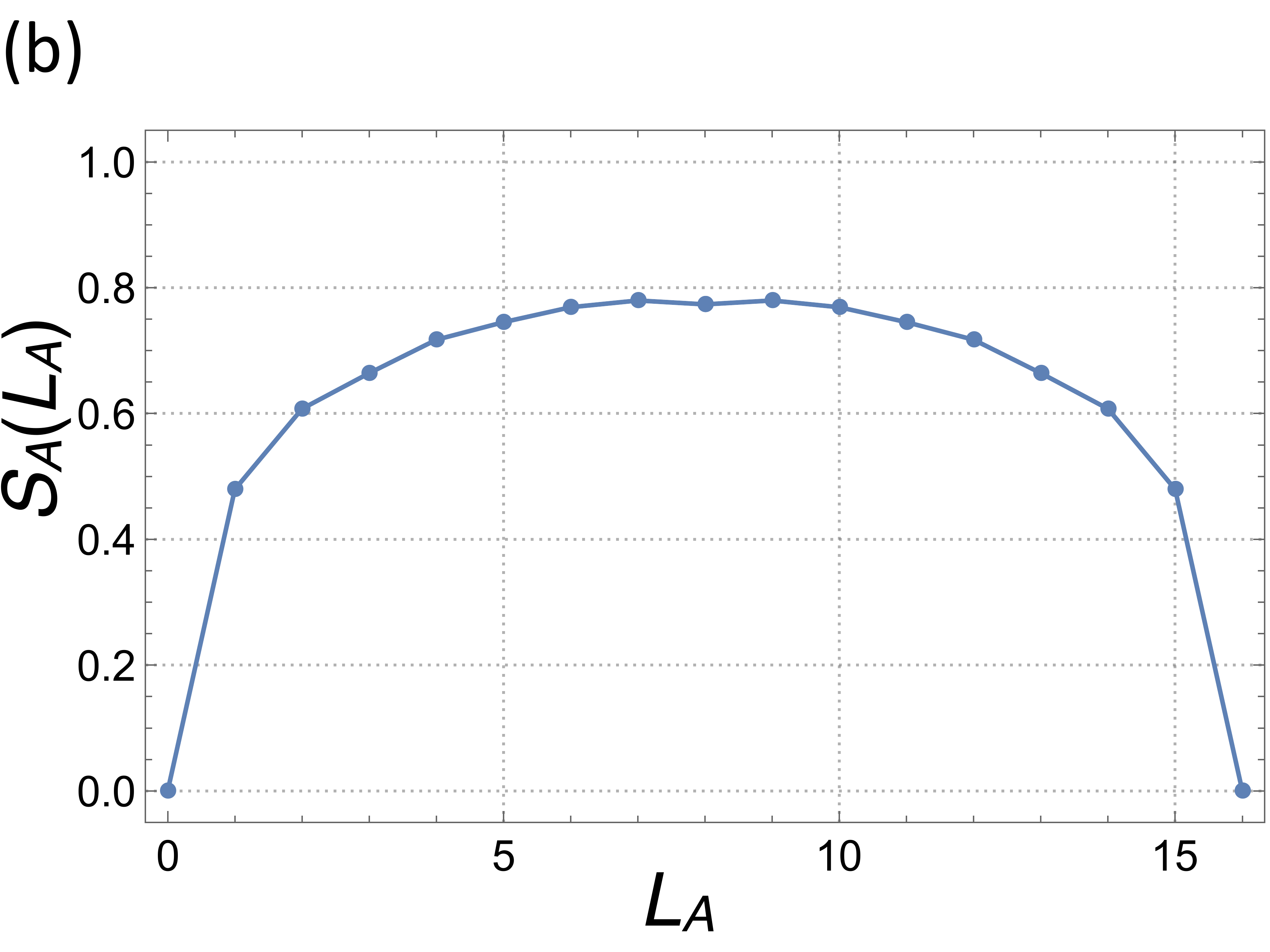}
     \includegraphics[width=.23\textwidth]{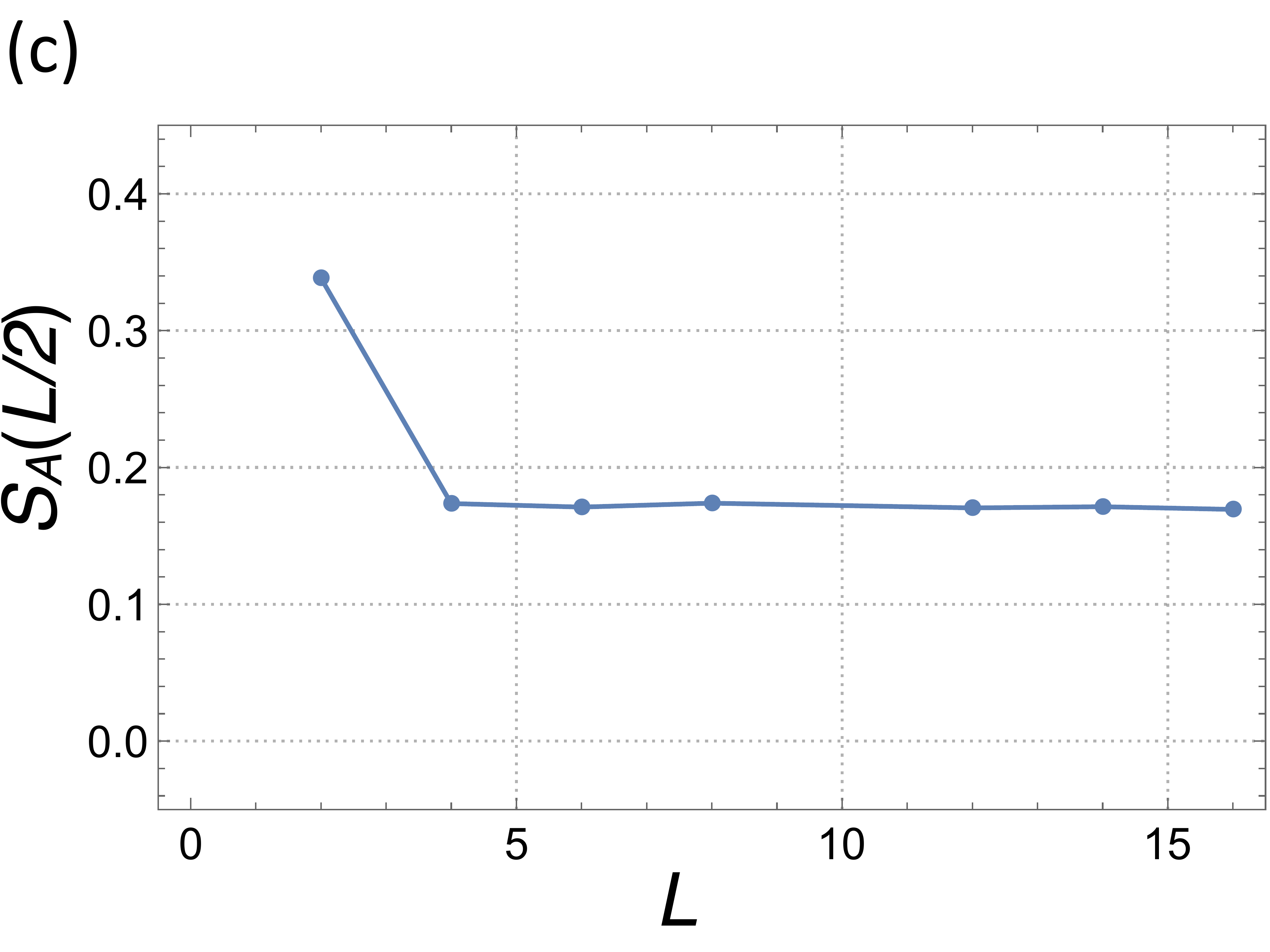}
     \includegraphics[width=.23\textwidth]{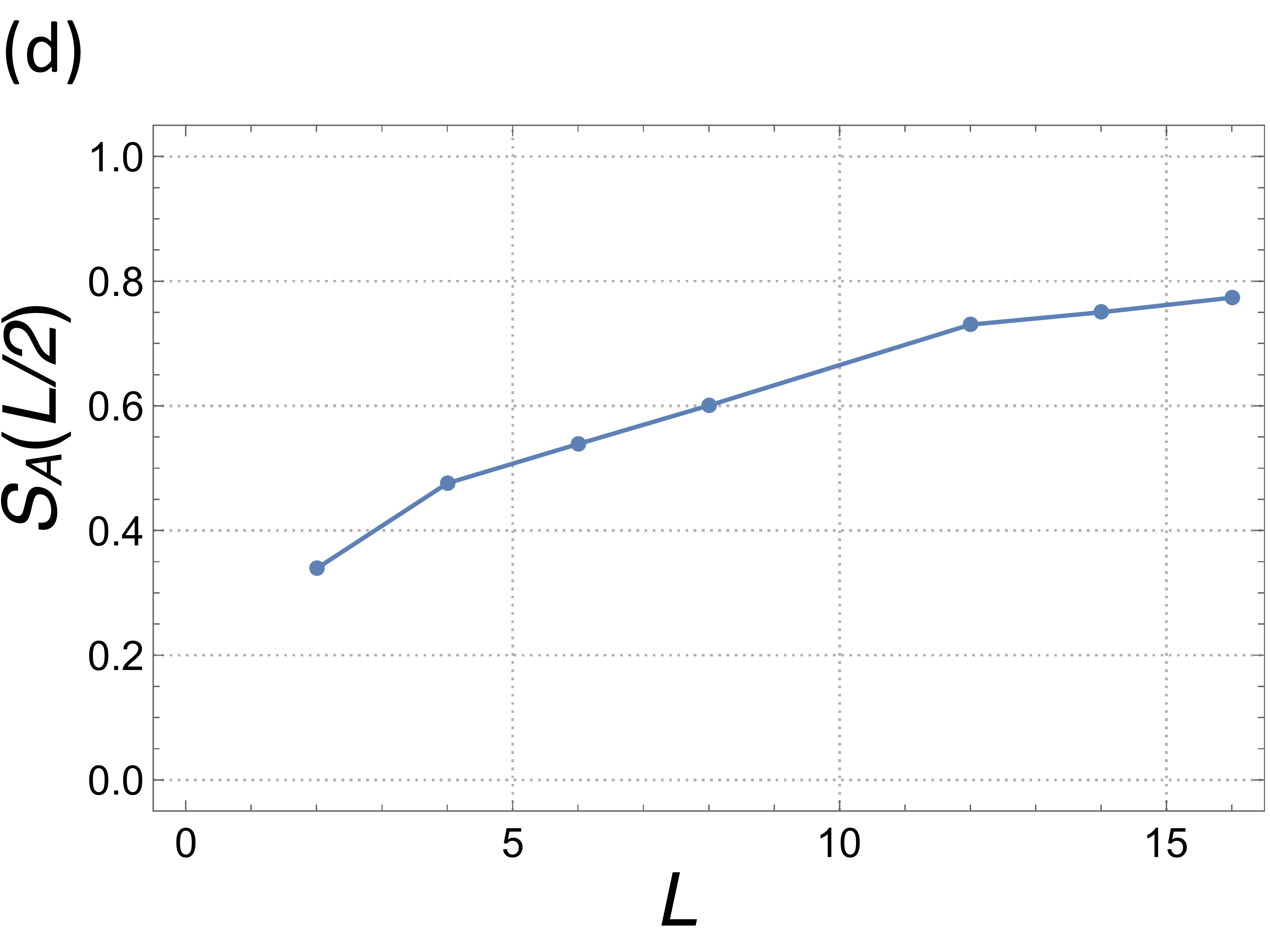}
     \includegraphics[width=.23\textwidth]{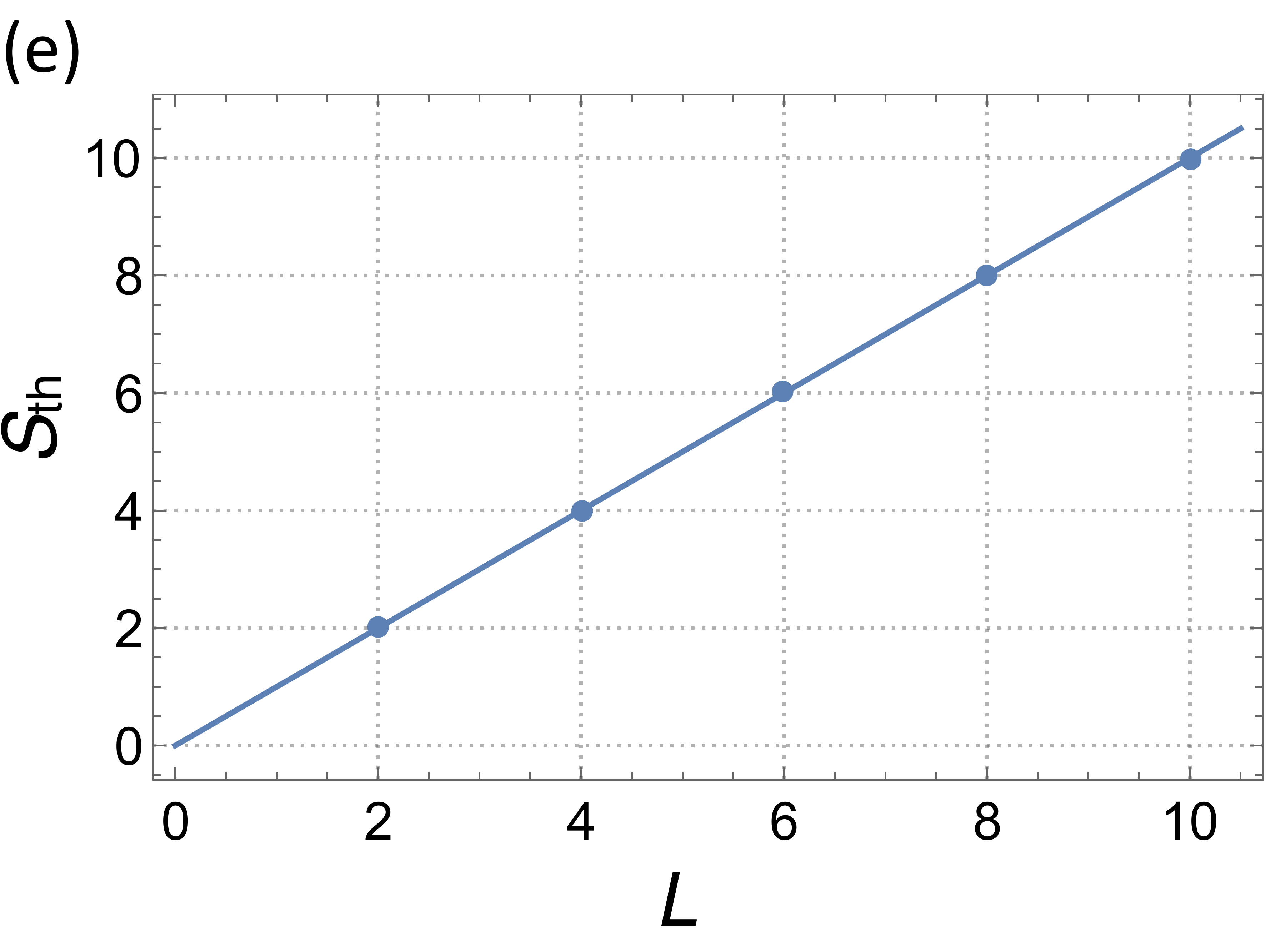}
     \includegraphics[width=.23\textwidth]{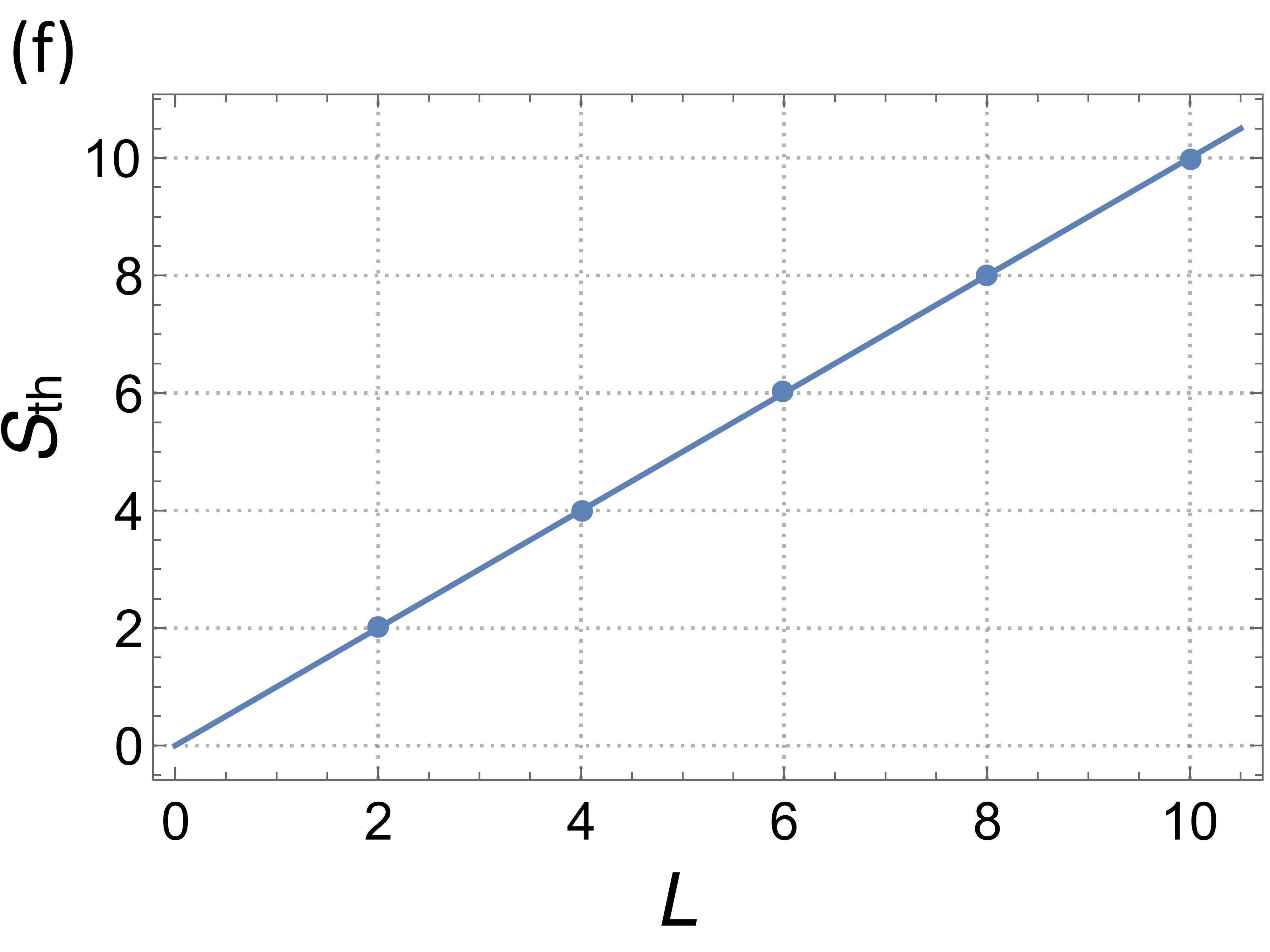}
     \caption{
     Results for model A1 (a,c,e) and model A2 (b,d,f).
     (a, b) The averaged steady-state value of the entanglement entropy for a system of size $L=16$ as a function of subsystem size $L_A$, for model A1 (a) and A2 (b), respectively.
     (a) There is almost no dependence of $S_A(L_A)$ on $L_A$, and this behavior strongly supports the area law of entanglement.
     (b) The behavior of $S_A(L_A)$ for model A2 is also consistent with an area law.
     (c, d) The averaged steady-state entanglement entropy for different system sizes $L$, while keeping $L_A / L = 1/2$ fixed, for model A1 (c) and A2 (d), respectively.
     (c) The half-system entanglement goes to a small constant as $L$ increases, again supporting the area law of entanglement.
     (d) The half-system entanglement appears to be sub-extensive in $L$.
     (e, f) The steady state value of thermal entropy as a function of system size for $L$ up to $10$ for A1 (e) and A2 (f), respectively.
     (e) The data points fits well on a straight line with slope $1$.
     (f) The data points fits well on a straight line with slope $1$.
  }
     \label{fig2}
 \end{figure}

 We implement numerical simulations of the quantum trajectories with system sizes up to $L=16$. For concreteness, we fix the initial state to be a product state with no entanglement,
 \begin{eqnarray}
    \ket{\psi_0} = \bigotimes_{i=1}^n \frac{1}{\sqrt{2}} \( \ket{0} + \ket{1} \).
 \end{eqnarray}
 We sample the unitary gates from a uniform Haar measure over the 2-qubit unitary group, and follow the quantum trajectory by recording the wavefunction.
 We keep evolving the system until the entanglement entropy saturates. For system sizes under consideration, we take the maximal duration of time to be $t = 2 \times 10^2$.
 We compute and average the entanglement entropy over an ensemble of $10^2$ samples.

 The results for $S_A(L_A, t \to \infty)$ versus $L_A$ are shown in Fig.~\ref{fig2}(a) for A1 and Fig.~\ref{fig2}(b) for A2.  From this data
 we can confidently conclude that the entanglement entropy saturates to an area law for {sufficiently} strong measurements in both models.
 This behavior is particularly easy to understand for model A1,
 where the measurements in a given time step project the qubits into a product state in the local $Z$ basis. The subsequent unitary transformations can only generate $O(1)$ entanglement between the two qubits upon which they act.
 In the next time step, this weakly entangled state is again projected onto a product state. Thus, the entanglement is at most $O(1)$ at all times for model A1 -  the area law will clearly persist in the thermodynamic limit.
 We say that the system is in a ``Zeno phase", in which the strong measurements have suppressed the entanglement, giving an area law.

 The results for $S_A(L/2, t \to \infty)$ versus $L$ are shown in Fig.~\ref{fig2}(c) for A1 and Fig.~\ref{fig2}(d) for A2.  While the data for model A1 is clearly consistent with an area law,
 there is a stronger system size dependence of the entanglement entropy in model A2.
 This difference is because the $\{P^{(2)}\}$ projectors are less effective in suppressing the entanglement entropy than the $\{P^{(1)}\}$ projectors, and sometimes even generate entanglement.
 But the magnitude of the entanglement entropy is small ($S_A < 1$) for system size up to $16$
 in both models, suggesting an area law scaling behavior.

 In Fig.~\ref{fig2}(e) and \ref{fig2}(f) we present results for the steady-state thermal entropy $S_{\rm th} (t\to\infty)$ up to system sizes of $L=10$, for models A1 and A2, respectively, obtained by computing the time dependence of the (mixed-state) density matrix, as described in Section~\ref{sec2}.
 {If we do not condition the unitaries on the measurement outcomes, we find that the thermal entropy is extensive in system size and in fact maximal.}
Thus, the system has ``infinite temperature'', as seen by the measuring environment.
 Nevertheless, by means of the successive measurements, the system manages to keeps its different parts weakly entangled, similar to zero temperature ground states of quantum many-body Hamiltonians. This striking difference between the entanglement and the thermal entropies is  characteristic of the unitary-projective circuit models.

\subsection{Quantum Phase Transition \label{sec3-B}}

\begin{figure}[t]
    \centering
    \includegraphics[width=.48\textwidth]{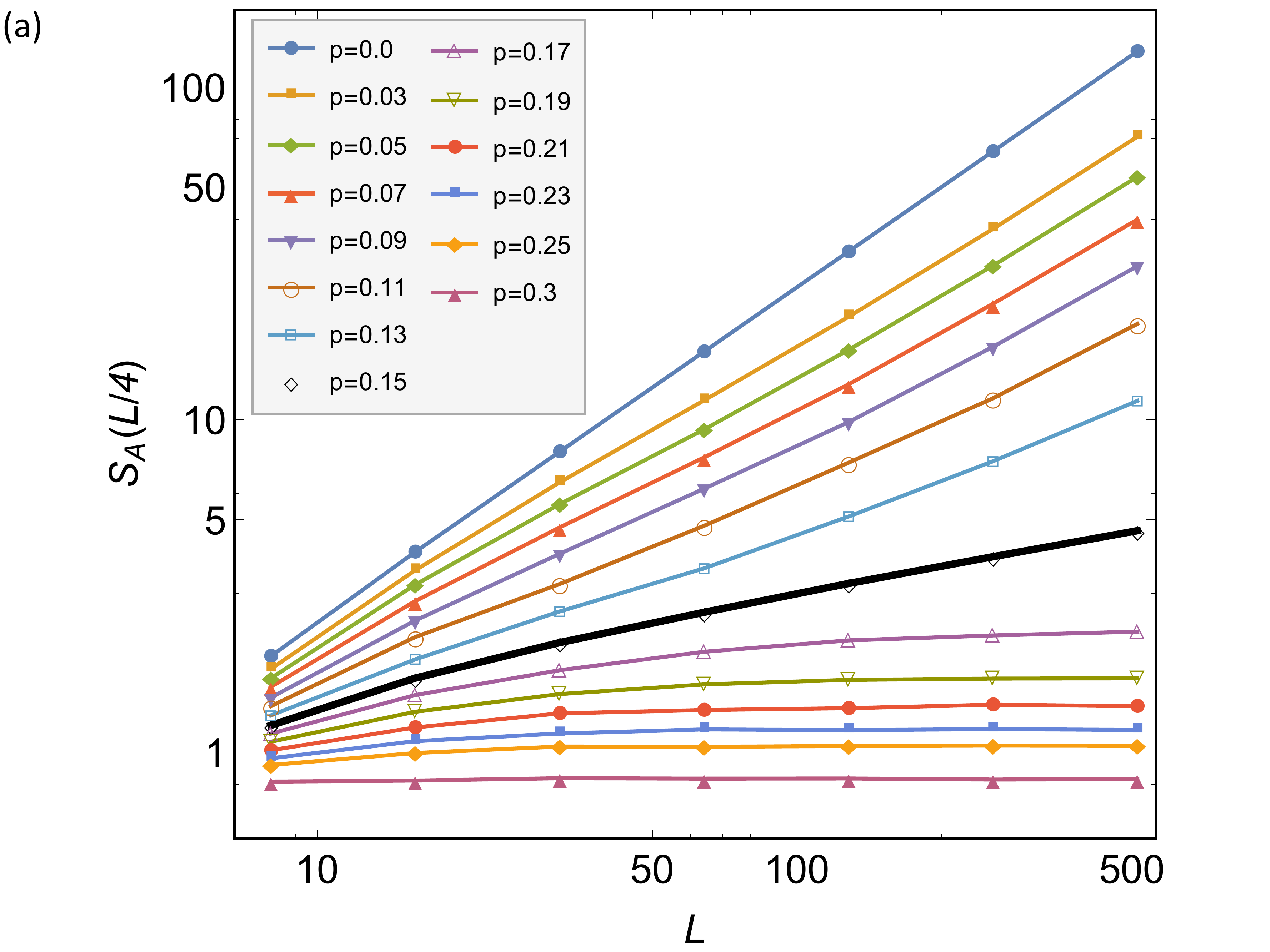}
    \includegraphics[width=.48\textwidth]{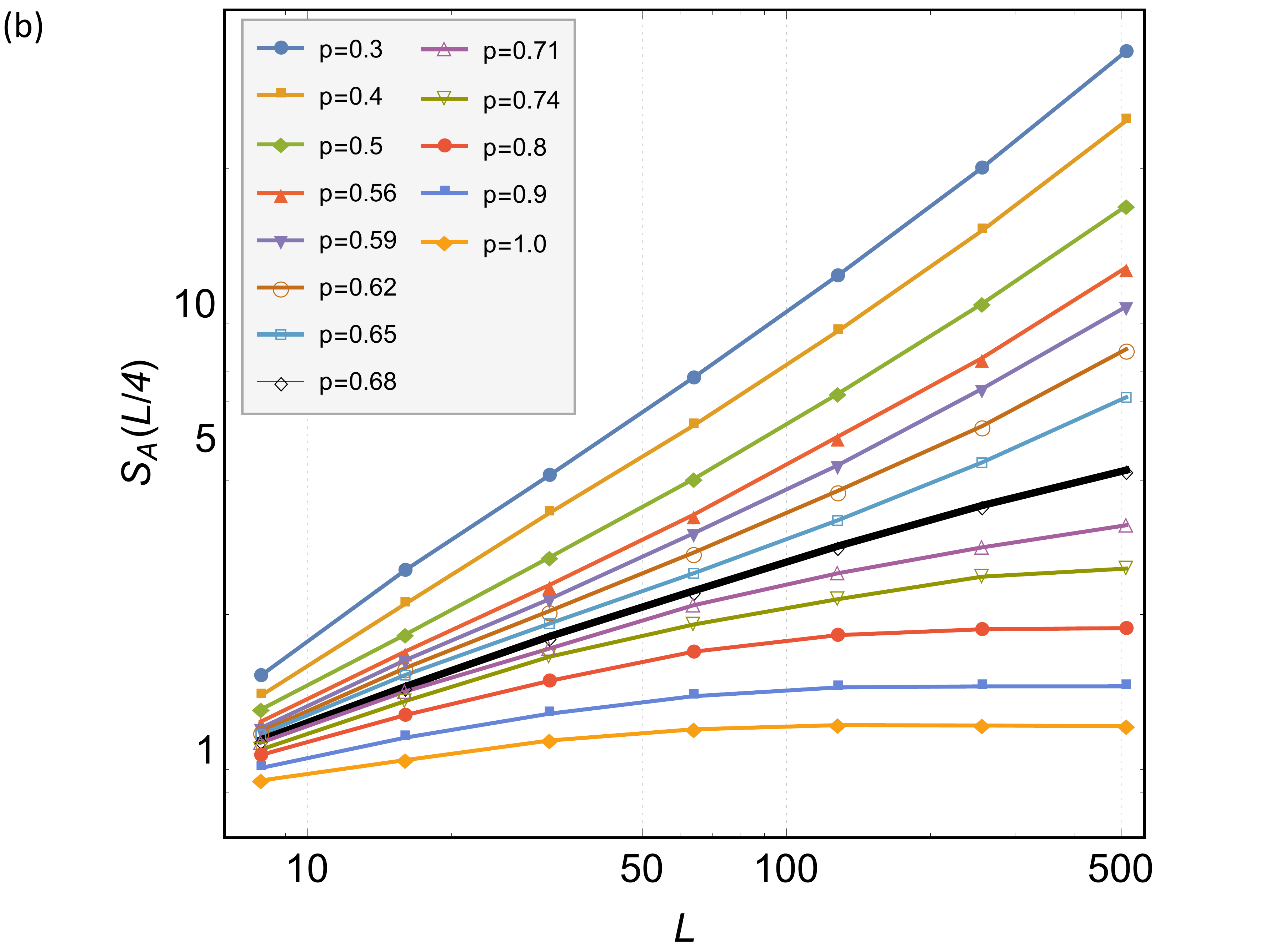}
    \caption{
    The averaged entanglement entropy for models B1 and B2 are shown in panels (a) and (b), respectively, as a function of system size, for different values of $p$, on a log-log scale. All the data is taken with subsystem size $L_A = L/4$.  In each figure one curve is highlighted with a thick line, corresponding to a critical value of $p=p_c$, that separates curves with $p<p_c$ that appear to asymptote to a straight line with slope $\approx 1$ at large $L$ (volume law), from the curves with $p>p_c$ which saturate to lines with slope $0$ at large $L$ (area law).  }
    \label{fig3}
\end{figure}

\begin{figure}[t]
    \centering
    \includegraphics[width=.48\textwidth]{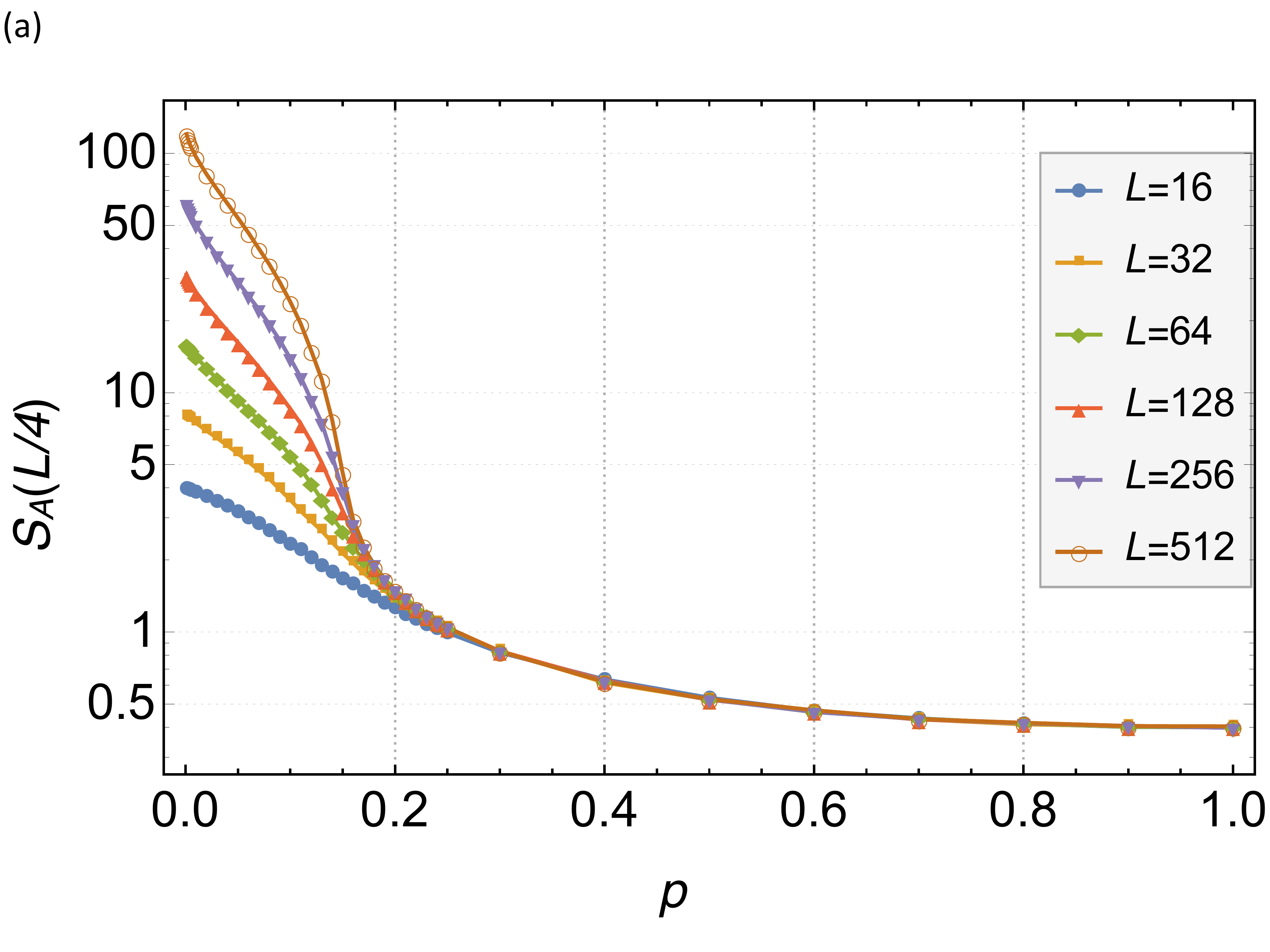}
    \includegraphics[width=.48\textwidth]{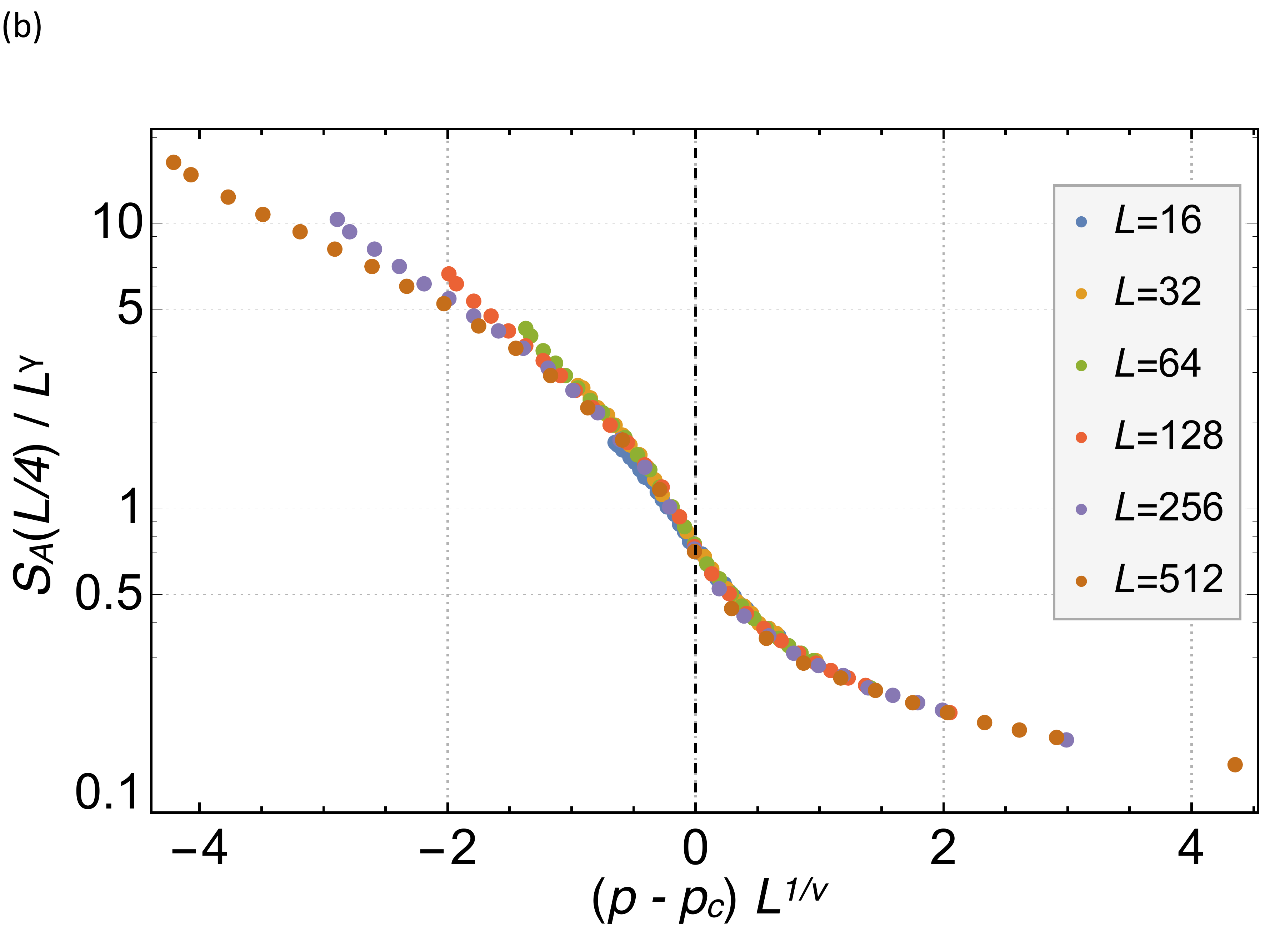}
    \caption{
    Results for model B1, plotted on a semi-log scale. In both panels, we take the subsystem size $L_A = L/4$.
    (a) The entanglement entropy as a function of $p$, for  different system sizes.
    (b) $S_A / L^{\gamma}$ versus $(p-p_c) L^{1/\nu}$, for $0.05 < p < 0.3$. We find $p_c = 0.15$, $\nu = 1.85$, $\gamma = 0.30$ for a best collapse.
 }
    \label{fig4}
\end{figure}

\begin{figure}[h]
    \centering
    \includegraphics[width=.48\textwidth]{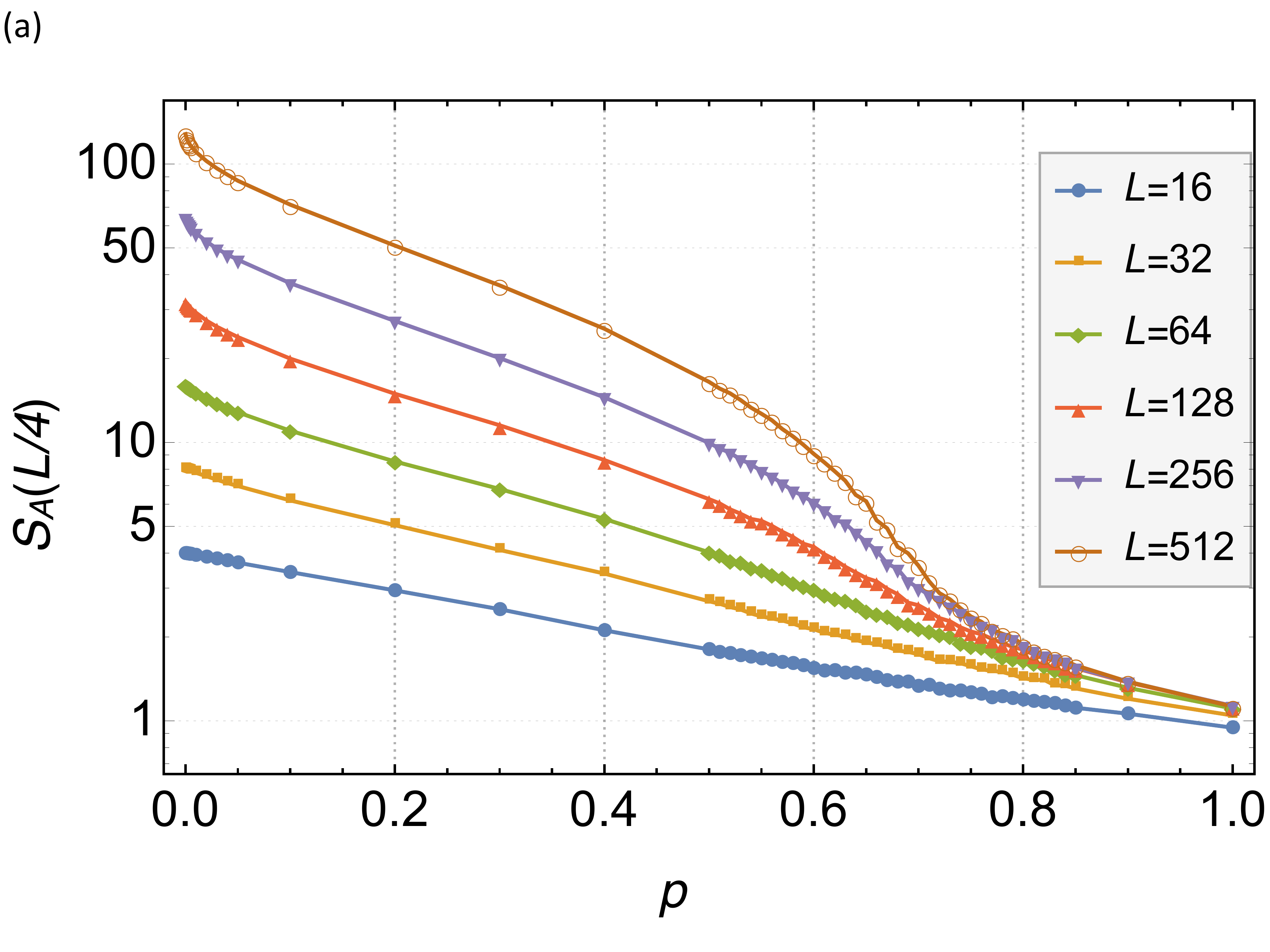}
    \includegraphics[width=.48\textwidth]{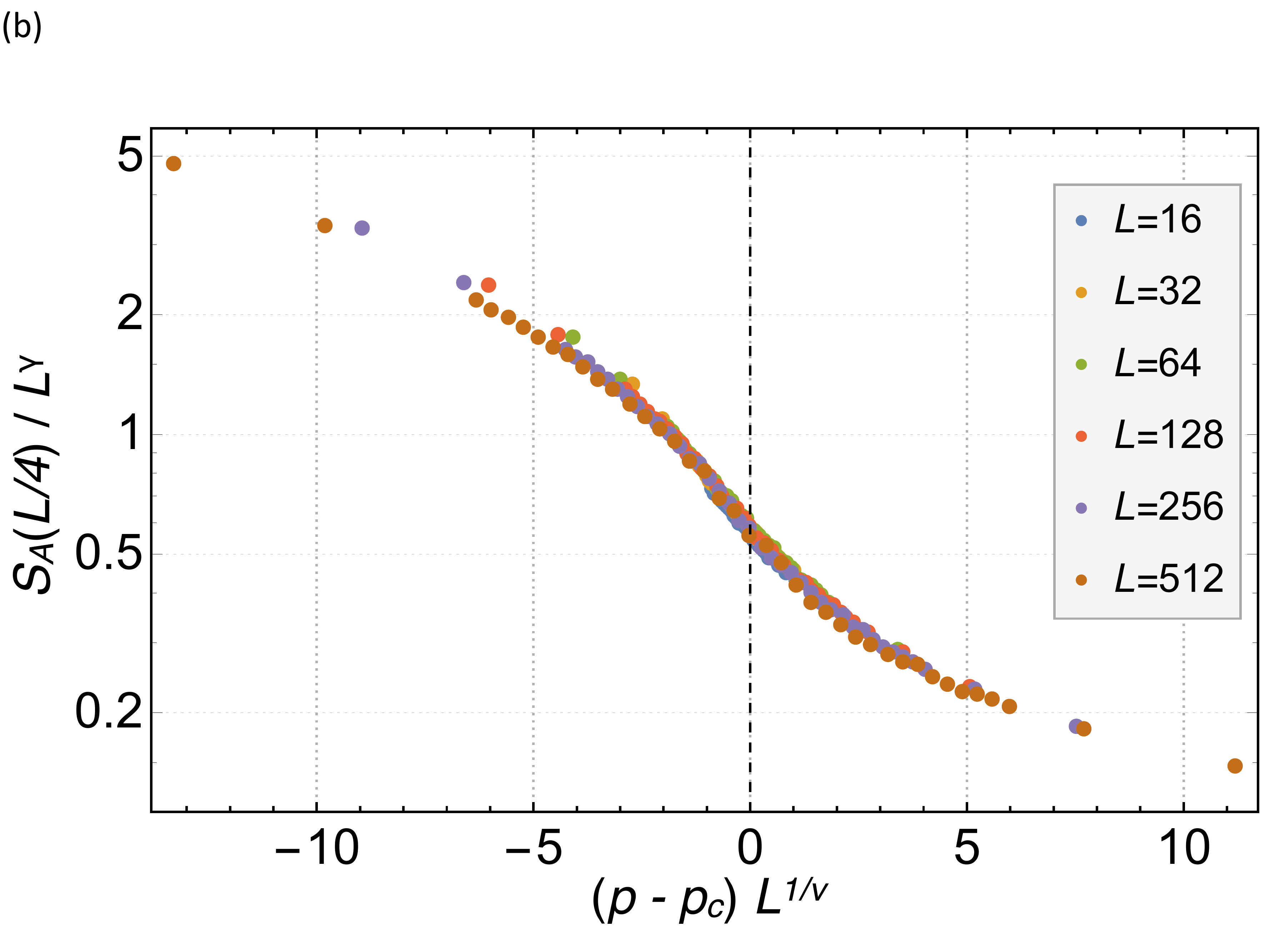}
    \caption{
    Results for model B2, plotted on a semi-log scale. In both panels, we take the subsystem size $L_A = L/4$.
   (a) The entanglement entropy as a function of $p$, for different system sizes.
   (b) $S_A / L^{\gamma}$ versus $(p-p_c) L^{1/\nu}$, for $0.3 < p < 1.0$. We find $p_c = 0.68$, $\nu = 1.75$, $\gamma = 0.33$ for a best collapse.
   }
    \label{fig5}
\end{figure}

{With the area law entanglement when $p\to 1$ established in subsection \ref{sec3-A} and volume law entanglement when $p\to 0$ established in \cite{nahum2017KPZ} },
we investigate the intermediate behavior when $p$ is tuned between these limits. In particular, we are interested in the possible existence of a critical rate $p_c > 0$, below which the average entanglement saturates to volume law.

To access larger system sizes, we turn to model B1 and B2, for which efficient classical algorithms are possible. We perform numerical simulations for systems with size $L$ up to $512$, and for a wide range of $p \in [0, 1]$. For conceptual simplicity we take the initial state to be a product state stabilized by the following stabilizers~\cite{nielsen2010qiqc},
\begin{eqnarray}
	\mathcal{S} = \{ X_1, X_2, \ldots, X_L \},
\end{eqnarray}
The long time steady state is expected to be independent of the initial state, but not so for the dynamical crossover that we explore in subsection~\ref{sec3-C}.

We sample the unitary gates from a uniform distribution over the 2-qubit Clifford group.
Since the projective measurements in Eq.~\eqref{eq:P1_00}--\eqref{eq:P2_1} are from the Pauli group, efficient simulation is
possible on a classical computer, as established by
the Gottesman-Knill theorem~\cite{gottesman9807heisenberg, aaronson0406chp, nielsen2010qiqc}.
We follow the quantum trajectories by keeping track of the stabilizers of the quantum state at each time step.  We note that for such stabilizer states the Renyi entropies (which are equal for all the Renyi indices)
can be readily computed~\cite{hamma2005bipartite, hamma2005entanglmeent, nahum2017KPZ}.
For system sizes under consideration, the entanglement entropy has already saturated within the duration of the simulation, $t =6 \times 10^2$.
Since the unitaries and the locations
and outcomes of the projective measurements are all random, we average the long-time entanglement entropy over an ensemble of quantum trajectories.
Typically we keep $10^2$ members in the ensemble.

The ensemble averaged steady state values of the entanglement entropy $S_A(L/4, t \to \infty)$ for the models B1 and B2 are shown in Fig.~\ref{fig3}(a) and Fig.~\ref{fig3}(b), respectively, where the subsystem size is $L_A = L/4$.  Here the entropy is plotted
versus system size (on a log-log scale) for different values of the projection probability $p$.
For both models we find evidence for a transition at a critical value $p=p_c$ (highlighted with a thick line),
with $p_c =  0.15$ for model B1 and $p_c = 0.68$ for model B2.  In both figures for $p< p_c$, the $\log S_A - \log L$ curves saturate to straight lines with slope $\approx 1$ for large $L$, suggesting a volume law.
For $p>p_c$, the curves seem to saturate to a horizontal line with zero slope for increasing $L$, i.e., the entanglement entropy is independent of $L$, suggesting an area law scaling behavior.
There is a rather clear distinction between the signs of the curvatures on either side of this putative transition.

To further probe this phase transition we first re-plot the entanglement entropy data versus
$p$ (on a semi-log scale) for the different systems sizes in Fig.~\ref{fig4}(a) for model B1 and Fig.~\ref{fig5}(a) for model B2, respectively.  And then we attempt a data collapse, fitting to the following standard finite-size scaling form near the critical point for the steady state entanglement $S_A(p, L_A)$,
\begin{eqnarray}
    S_A(p, c L) = L ^\gamma F \( (p-p_c) L^{1/\nu} \),
\end{eqnarray}
where the values of $p_c$ are taken from Fig.~\ref{fig3}, and $c=L_A/L \le 1/2$ is a finite constant, fixed to be $c=1/4$ in our numerics \footnote{We note that in Refs. \cite{Vasseur2018, nahum2018hybrid}, a different scaling form for entanglement entropy, $S_A(p, L_A) = a(p) \log L_A + b(p) L_A$ was used for data collapse.}.
Restricting the values of $p$ to be close to criticality,
$0.05<p<0.3$ and $0.3 < p < 1.0$ for the two cases, we re-plot the data
as $S_A / L^{\gamma}$ versus $(p-p_c) L^{1/\nu}$, and choose the critical exponents $\gamma$ and $\nu$ to get the best collapse.  For model B1 we find that $\nu = 1.85$ and $\gamma = 0.30$
give the best fit as shown in Fig.~\ref{fig4}(b), while the data collapse for model B2 with $\nu = 1.75$ and $\gamma = 0.33$ is shown in Fig.~\ref{fig5}(b).

The quality of the data collapse, and the
closeness of the critical exponents between the two models, lend strong support for the existence
of a continuous quantum phase transition.  Below $p_c$, there is a stable ``weak measurement'' phase in which the system saturates to volume law entanglement entropy. Above $p_c$, the system is in the ``strong measurement'' or Zeno phase, where the entanglement entropy saturates to an area law.
Right at the transition the entanglement entropy is growing with $L$ but sub-extensive,
$S_A(p_c, c L) \sim L^\gamma$ with $\gamma \approx 1/3$.  Upon approaching the critical point there is a diverging correlation
length, $\xi \sim |p-p_c|^{-\nu}$, as deduced from the finite-size scaling collapse.

Once the critical exponents are known, the behavior of the scaling function, $F(x)$,
at large argument $|x| \gg 1$ can be deduced by matching on to the
volume and area law behaviors expected in the weak and strong measurement phases
on either side of the transition.  To wit, requiring  $S_A \sim L_A^1$ for  $p \lesssim p_c$, and $S_A \sim L_A^0$ for $p \gtrsim p_c$, dictates the large $x$ behavior of the scaling functions,
\begin{eqnarray}
	F(x) \sim \begin{cases}
		|x|^{(1-\gamma)\nu}, \,x \to -\infty,\\
		x^{-\gamma \nu}, \quad \ \ x \to +\infty .
	\end{cases}
\end{eqnarray}
In turn, this implies,
\begin{eqnarray}
    \lim_{L \to \infty} L^{-1} S_A(p, c L) &\sim& |p-p_c|^{(1-\gamma)\nu}, \text{ for } p < p_c,\\
    \lim_{L \to \infty} S_A(p, c L) &\sim& (p-p_c)^{-\gamma\nu}, \text{ for } p > p_c.
\end{eqnarray}
Notice that the coefficient of the volume law of the entanglement entropy (i.e. the entanglement entropy density) vanishes continuously upon approaching the critical point ($p \rightarrow p_c^-$) with exponent $(1-\gamma)\nu \approx 4/3$.   As such, the entanglement entropy density can serve as an ``order parameter" for the ``weak measurement phase", analogous to the magnetization in an Ising transition non-zero for $T < T_c$.   On the other hand, the value of the area law entanglement entropy in the strong measurement Zeno phase diverges upon approaching the critical point,
with exponent $\gamma \nu \approx 2/3$.

Despite the different set of projectors in models B1 and B2, their measured critical exponents (both $\nu$ and $\gamma$)
are very close to one another, suggesting that they belong to the same universality class.
We also suspect that the choice of Clifford unitaries is not so important, and that
Haar random (or other) unitaries would lead to a transition with the same universal properties.

\subsection{Critical Dynamic Scaling}
\label{sec3-C}

\begin{figure}[h]
    \centering
    \includegraphics[width=.48\textwidth]{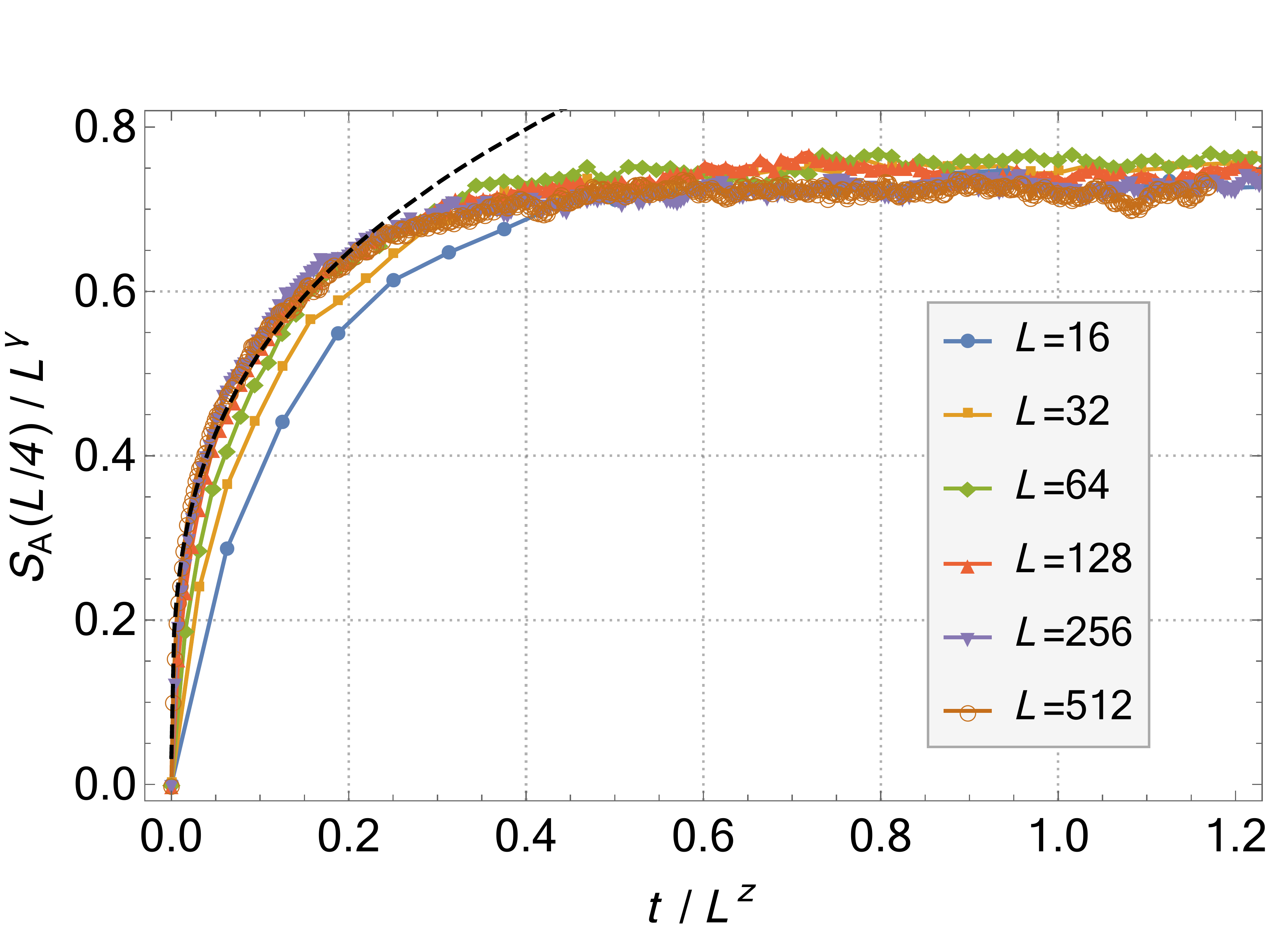}
    \caption{
    $S_A L^{-\gamma}$ versus $t L^{-z}$ (where $z=1$) at criticality, $p=p_c$, in model B1. The entanglement entropy is computed for a subsystem of size $L_A = L/4$.
    The dashed line shows the funtion $f(x) \sim x^{\gamma/z}$, where $x = t L^{-z}$, which matches the data collapse well as $x \to 0$.
    }
    \label{fig6}
\end{figure}

Previously we have been focusing on the {steady-state} behavior of the entanglement entropy as $t \to \infty$. At the transition, we found $S_A \sim L^\gamma$, consistent with the diverging length scale, $\xi \sim |p-p_c|^{-\nu}$ near criticality.
By analogy with conventional dynamical critical phenomena, we expect
an associated diverging time scale, $\tau \sim \xi^z$, with dynamic exponent $z$, corresponding
to critical slowing down.
Taking the dynamic scaling into account, we can generalize the finite size scaling form to include time,
\begin{eqnarray}
    \label{eq:scaling}
    S_A(p, c L, t) = L^\gamma F\((p-p_c) L^{1/\nu}; t L^{-z}\),
\end{eqnarray}
where we are assuming a initial product state at time $t=0$.
At criticality this reduces to,
\begin{eqnarray}
    \label{eq:scaling_critical}
    S_A(p_c, c L, t)  = L^{\gamma} f \(t L^{-z}\),
\end{eqnarray}
where $f(x) \equiv F(0, x)$. In Fig.~\ref{fig6}, we plot $S_A(p_c, L/4, t) L^{-\gamma}$ as a function of $t L^{-z}$, with $z=1$ in model B1. The quality of the data collapse indicates that
dynamic scaling is operative, with a dynamic critical exponent $z \approx 1$.
As expected the scaling function $f(x \rightarrow \infty) \sim \text{const}$ (see Fig.~\ref{fig6}), recovering the steady-state behavior, $S_A(p_c, c L) \sim L^\gamma$.
In the opposite limit, $x=t L^{-z} \ll 1$, the $L$-dependence in Eq.~\eqref{eq:scaling_critical} must cancel out, since the initial growth of entanglement of an unentangled initial state is strictly local, and insensitive to the system size. Thus, the
scaling function must vary as $f(x \rightarrow 0) \sim x^{\gamma/z}$, implying sub-linear power-law entanglement growth at criticality,
$S_A(p_c,t \ll L^z) \sim t^{\gamma/z}$.  The exponent $\gamma/z \approx 1/3$ is consistent with the measured scaling function for short times in Fig.~\ref{fig6}.

In the thermodynamic limit $L\to\infty$, Eq.~\eqref{eq:scaling} can be re-expressed as
\begin{equation}
    S_A(p, c L \to \infty,t) = t^{\gamma/z} G\((p-p_c)t^{1/z\nu}\).
\end{equation}
We know that in the $p\to 0$ limit, the entanglement grows linearly in time before saturating to volume law,
and we expect this to be the case quite generally in the weak measurement phase with $p<p_c$.
This implies that the scaling function $G(y) \sim |y|^{\nu(\gamma-z)}$ for $y \rightarrow - \infty$
and that,
\begin{eqnarray}
    S_A (p\lesssim p_c, cL\to \infty, t) \sim |p-p_c|^{\nu(z-\gamma)}t,
\end{eqnarray}
so that the entanglement velocity vanishes upon approaching
criticality, $v_E \sim |p-p_c|^{\nu(z-\gamma)}$, with an exponent $\nu(z-\gamma) \approx 4/3$,
a signature of critical slowing down.

When $p$ is below $p_c$, the time scale $\tau \sim \xi^z$ is finite, and for $L_A \gg v_E \tau$, the growth of entanglement entropy can be divided into three regimes,
\begin{eqnarray}
    S_A(p<p_c, L_A \gg v_E \tau, t) \sim \begin{cases}
        t^{\gamma/z}, & \text{ if } t \ll \tau,\\
        v_E t, & \text { if } \tau \ll t \ll \tau_\ast,\\
        L_A, & \text { if } t \gg \tau_\ast,
    \end{cases}
\end{eqnarray}
where $\tau_\ast = L_A / v_E$.


\section{Discussion \label{sec5}}
In this paper we explored the competition between unitary time evolution and projective measurements in a 1d quantum circuit with chaotic dynamics.  We performed extensive numerics following the many-body quantum trajectories for both Haar and Cifford random circuits.
Upon varying the measurement rate, we found that the steady states are either in the volume law entangled phase (weak measurement) or the ``quantum Zeno phase" with area law entanglement entropy (strong measurement). For the Clifford circuit we were able to access a continuous dynamical quantum entanglement transition separating these two phases.  For systems up to 512 qubits, our data could be collapsed using standard finite size scaling, allowing critical exponents to be extracted.  The transition exhibited critical slowing down, with a dynamic exponent given by $z \approx 1$.   This many-body entanglement transition was driven by a competition between the entangling tendencies of unitary time evolution and the disentangling tendencies of projective measurements.

In the past there has been extensive work exploring possible wave function entanglement entropy transitions~\cite{Chen2015, Vasseur2018}, the best known example being the
many-body localization (MBL) transition~\cite{Kjall2014,Luitz2015, Vosk2015, Potter2015, Serbyn2015, Zhang2016, Dumitrescu2017,Khemani2017}.  In the localized phase the highly excited eigenstates have an area law entanglement entropy~\cite{Pal2010,Bauer2013,Nandkishore2015}, while the
excited eigenstates in the delocalized phase satisfy the eigenstate thermalization hypothesis with volume law
entanglement entropy~\cite{Deutsch1991,Srednicki1994}.
Quenched disorder drives the MBL transition, while
 the area law phase in our circuit model is driven by strong measurements.
Moreover, the critical behavior near the MBL transition is quite unusual,
exhibiting activated scaling upon approaching the
transition from the thermal side with dynamics slowing due to the Griffiths effect~\cite{Vosk2015,Potter2015,Griffiths1969} and the entanglement spreads sub-ballistically as $t^{1/z}$~\cite{Agarwal2015,Torres2015,Luitz2016} with the dynamic exponent $z$ diverging at the critical point~\cite{Prosen2008,Bardarson2012,Serbyn2013, Vosk2015,Potter2015}.
In contrast, the critical behavior at the measurement driven entanglement entropy transition in
the hybrid unitary/measurement circuit satisfies conventional dynamic scaling
with $z \approx 1$, rather than
activated scaling.   Moreover, right at the MBL transition the entanglement entropy is still a volume law~\cite{Vosk2015,Potter2015}, with the
coefficient of the volume law jumping discontinuously to zero upon crossing the
transition~\cite{Vosk2015,Dumitrescu2017,Grover2014}.
In contrast, right at the measurement driven transition in our hybrid
circuit, a sub-linear size-scaling of the entanglement entropy is found, and the coefficient of the volume law entanglement vanishes continuously upon approaching the transition
from the weak measurement phase.

While our numerics established the existence and critical properties of a 1d measurement driven entanglement transition, much work remains.
It would be interesting to explore the dimensionality dependence of the critical properties,
and to find an analytic treatment.
The sub-linear size-scaling of the entanglement entropy
right at the transition suggests that finding a possible dual holographic description
could be intriguing and informative.

\emph{Note added}:
We would like to draw the reader's attention to two related parallel works --- by Chan, Nandkishore, Pretko, and Smith~\cite{nandkishore2018hybrid}; and by Skinner, Ruhman, and Nahum~\cite{nahum2018hybrid} --- to appear in the same arXiv posting.

\section*{ACKNOWLEDGMENTS}
We thank Ehud Altman, Leon Balents, Anushya Chandran, Michael Freedman, Daniel Gottesman, Timothy Hsieh, Adam Nahum, Mark Rudner, and Tianci Zhou for fruitful discussions.
X. C. was supported by a postdoctoral fellowship from the Gordon and Betty Moore
Foundation, under the EPiQS initiative, Grant GBMF4304, at the Kavli Institute for Theoretical
Physics.
M. P. A. F. is grateful to the Heising-Simons Foundation for support,
to the National Science Foundation for support under Grant No. DMR-1404230,
and to the Caltech Institute of Quantum Information and Matter, an NSF Physics Frontiers Center
with support of the Gordon and Betty Moore Foundation.
Computational resources were provided by the Center for Scientific Computing from the CNSI, MRL: an NSF MRSEC (DMR-1720256) and NSF CNS-1725797.

\bibliography{refs}

\begin{thebibliography}{61}%
\makeatletter
\providecommand \@ifxundefined [1]{%
 \@ifx{#1\undefined}
}%
\providecommand \@ifnum [1]{%
 \ifnum #1\expandafter \@firstoftwo
 \else \expandafter \@secondoftwo
 \fi
}%
\providecommand \@ifx [1]{%
 \ifx #1\expandafter \@firstoftwo
 \else \expandafter \@secondoftwo
 \fi
}%
\providecommand \natexlab [1]{#1}%
\providecommand \enquote  [1]{``#1''}%
\providecommand \bibnamefont  [1]{#1}%
\providecommand \bibfnamefont [1]{#1}%
\providecommand \citenamefont [1]{#1}%
\providecommand \href@noop [0]{\@secondoftwo}%
\providecommand \href [0]{\begingroup \@sanitize@url \@href}%
\providecommand \@href[1]{\@@startlink{#1}\@@href}%
\providecommand \@@href[1]{\endgroup#1\@@endlink}%
\providecommand \@sanitize@url [0]{\catcode `\\12\catcode `\$12\catcode
  `\&12\catcode `\#12\catcode `\^12\catcode `\_12\catcode `\%12\relax}%
\providecommand \@@startlink[1]{}%
\providecommand \@@endlink[0]{}%
\providecommand \url  [0]{\begingroup\@sanitize@url \@url }%
\providecommand \@url [1]{\endgroup\@href {#1}{\urlprefix }}%
\providecommand \urlprefix  [0]{URL }%
\providecommand \Eprint [0]{\href }%
\providecommand \doibase [0]{http://dx.doi.org/}%
\providecommand \selectlanguage [0]{\@gobble}%
\providecommand \bibinfo  [0]{\@secondoftwo}%
\providecommand \bibfield  [0]{\@secondoftwo}%
\providecommand \translation [1]{[#1]}%
\providecommand \BibitemOpen [0]{}%
\providecommand \bibitemStop [0]{}%
\providecommand \bibitemNoStop [0]{.\EOS\space}%
\providecommand \EOS [0]{\spacefactor3000\relax}%
\providecommand \BibitemShut  [1]{\csname bibitem#1\endcsname}%
\let\auto@bib@innerbib\@empty
\bibitem [{\citenamefont {{Kitaev}}\ and\ \citenamefont
  {{Preskill}}(2006)}]{Kitaev2006}%
  \BibitemOpen
  \bibfield  {author} {\bibinfo {author} {\bibfnamefont {A.}~\bibnamefont
  {{Kitaev}}}\ and\ \bibinfo {author} {\bibfnamefont {J.}~\bibnamefont
  {{Preskill}}},\ }\href {\doibase 10.1103/PhysRevLett.96.110404} {\bibfield
  {journal} {\bibinfo  {journal} {Physical Review Letters}\ }\textbf {\bibinfo
  {volume} {96}},\ \bibinfo {eid} {110404} (\bibinfo {year} {2006})},\ \Eprint
  {http://arxiv.org/abs/hep-th/0510092} {hep-th/0510092} \BibitemShut {NoStop}%
\bibitem [{\citenamefont {{Levin}}\ and\ \citenamefont
  {{Wen}}(2006)}]{Levin2006}%
  \BibitemOpen
  \bibfield  {author} {\bibinfo {author} {\bibfnamefont {M.}~\bibnamefont
  {{Levin}}}\ and\ \bibinfo {author} {\bibfnamefont {X.-G.}\ \bibnamefont
  {{Wen}}},\ }\href {\doibase 10.1103/PhysRevLett.96.110405} {\bibfield
  {journal} {\bibinfo  {journal} {Physical Review Letters}\ }\textbf {\bibinfo
  {volume} {96}},\ \bibinfo {eid} {110405} (\bibinfo {year} {2006})},\ \Eprint
  {http://arxiv.org/abs/cond-mat/0510613} {cond-mat/0510613} \BibitemShut
  {NoStop}%
\bibitem [{\citenamefont {{Calabrese}}\ and\ \citenamefont
  {{Cardy}}(2004)}]{Calabrese2004}%
  \BibitemOpen
  \bibfield  {author} {\bibinfo {author} {\bibfnamefont {P.}~\bibnamefont
  {{Calabrese}}}\ and\ \bibinfo {author} {\bibfnamefont {J.}~\bibnamefont
  {{Cardy}}},\ }\href {\doibase 10.1088/1742-5468/2004/06/P06002} {\bibfield
  {journal} {\bibinfo  {journal} {Journal of Statistical Mechanics: Theory and
  Experiment}\ }\textbf {\bibinfo {volume} {6}},\ \bibinfo {pages} {06002}
  (\bibinfo {year} {2004})},\ \Eprint {http://arxiv.org/abs/hep-th/0405152}
  {hep-th/0405152} \BibitemShut {NoStop}%
\bibitem [{\citenamefont {{Ryu}}\ and\ \citenamefont
  {{Takayanagi}}(2006)}]{Ryu2006}%
  \BibitemOpen
  \bibfield  {author} {\bibinfo {author} {\bibfnamefont {S.}~\bibnamefont
  {{Ryu}}}\ and\ \bibinfo {author} {\bibfnamefont {T.}~\bibnamefont
  {{Takayanagi}}},\ }\href {\doibase 10.1103/PhysRevLett.96.181602} {\bibfield
  {journal} {\bibinfo  {journal} {Physical Review Letters}\ }\textbf {\bibinfo
  {volume} {96}},\ \bibinfo {eid} {181602} (\bibinfo {year} {2006})},\ \Eprint
  {http://arxiv.org/abs/hep-th/0603001} {hep-th/0603001} \BibitemShut {NoStop}%
\bibitem [{\citenamefont {{Casini}}\ and\ \citenamefont
  {{Huerta}}(2012)}]{Casini2012}%
  \BibitemOpen
  \bibfield  {author} {\bibinfo {author} {\bibfnamefont {H.}~\bibnamefont
  {{Casini}}}\ and\ \bibinfo {author} {\bibfnamefont {M.}~\bibnamefont
  {{Huerta}}},\ }\href {\doibase 10.1103/PhysRevD.85.125016} {\bibfield
  {journal} {\bibinfo  {journal} {\prd}\ }\textbf {\bibinfo {volume} {85}},\
  \bibinfo {eid} {125016} (\bibinfo {year} {2012})},\ \Eprint
  {http://arxiv.org/abs/1202.5650} {arXiv:1202.5650 [hep-th]} \BibitemShut
  {NoStop}%
\bibitem [{\citenamefont {{Page}}(1993)}]{Page1993}%
  \BibitemOpen
  \bibfield  {author} {\bibinfo {author} {\bibfnamefont {D.~N.}\ \bibnamefont
  {{Page}}},\ }\href {\doibase 10.1103/PhysRevLett.71.1291} {\bibfield
  {journal} {\bibinfo  {journal} {Physical Review Letters}\ }\textbf {\bibinfo
  {volume} {71}},\ \bibinfo {pages} {1291} (\bibinfo {year} {1993})},\ \Eprint
  {http://arxiv.org/abs/gr-qc/9305007} {gr-qc/9305007} \BibitemShut {NoStop}%
\bibitem [{\citenamefont {{Goldstein}}\ \emph {et~al.}(2006)\citenamefont
  {{Goldstein}}, \citenamefont {{Lebowitz}}, \citenamefont {{Tumulka}},\ and\
  \citenamefont {{Zangh{\`i}}}}]{Goldstein2006}%
  \BibitemOpen
  \bibfield  {author} {\bibinfo {author} {\bibfnamefont {S.}~\bibnamefont
  {{Goldstein}}}, \bibinfo {author} {\bibfnamefont {J.~L.}\ \bibnamefont
  {{Lebowitz}}}, \bibinfo {author} {\bibfnamefont {R.}~\bibnamefont
  {{Tumulka}}}, \ and\ \bibinfo {author} {\bibfnamefont {N.}~\bibnamefont
  {{Zangh{\`i}}}},\ }\href {\doibase 10.1103/PhysRevLett.96.050403} {\bibfield
  {journal} {\bibinfo  {journal} {Physical Review Letters}\ }\textbf {\bibinfo
  {volume} {96}},\ \bibinfo {eid} {050403} (\bibinfo {year} {2006})},\ \Eprint
  {http://arxiv.org/abs/cond-mat/0511091} {cond-mat/0511091} \BibitemShut
  {NoStop}%
\bibitem [{\citenamefont {{Calabrese}}\ and\ \citenamefont
  {{Cardy}}(2005)}]{Calabrese_2005}%
  \BibitemOpen
  \bibfield  {author} {\bibinfo {author} {\bibfnamefont {P.}~\bibnamefont
  {{Calabrese}}}\ and\ \bibinfo {author} {\bibfnamefont {J.}~\bibnamefont
  {{Cardy}}},\ }\href {\doibase 10.1088/1742-5468/2005/04/P04010} {\bibfield
  {journal} {\bibinfo  {journal} {Journal of Statistical Mechanics: Theory and
  Experiment}\ }\textbf {\bibinfo {volume} {4}},\ \bibinfo {pages} {04010}
  (\bibinfo {year} {2005})},\ \Eprint {http://arxiv.org/abs/cond-mat/0503393}
  {cond-mat/0503393} \BibitemShut {NoStop}%
\bibitem [{\citenamefont {{Calabrese}}\ and\ \citenamefont
  {{Cardy}}(2007)}]{Calabrese_2007}%
  \BibitemOpen
  \bibfield  {author} {\bibinfo {author} {\bibfnamefont {P.}~\bibnamefont
  {{Calabrese}}}\ and\ \bibinfo {author} {\bibfnamefont {J.}~\bibnamefont
  {{Cardy}}},\ }\href {\doibase 10.1088/1742-5468/2007/06/P06008} {\bibfield
  {journal} {\bibinfo  {journal} {Journal of Statistical Mechanics: Theory and
  Experiment}\ }\textbf {\bibinfo {volume} {6}},\ \bibinfo {pages} {06008}
  (\bibinfo {year} {2007})},\ \Eprint {http://arxiv.org/abs/0704.1880}
  {arXiv:0704.1880 [cond-mat.stat-mech]} \BibitemShut {NoStop}%
\bibitem [{\citenamefont {{Kim}}\ and\ \citenamefont {{Huse}}(2013)}]{Kim2013}%
  \BibitemOpen
  \bibfield  {author} {\bibinfo {author} {\bibfnamefont {H.}~\bibnamefont
  {{Kim}}}\ and\ \bibinfo {author} {\bibfnamefont {D.~A.}\ \bibnamefont
  {{Huse}}},\ }\href {\doibase 10.1103/PhysRevLett.111.127205} {\bibfield
  {journal} {\bibinfo  {journal} {Physical Review Letters}\ }\textbf {\bibinfo
  {volume} {111}},\ \bibinfo {eid} {127205} (\bibinfo {year} {2013})},\ \Eprint
  {http://arxiv.org/abs/1306.4306} {arXiv:1306.4306 [quant-ph]} \BibitemShut
  {NoStop}%
\bibitem [{\citenamefont {{Mezei}}\ and\ \citenamefont
  {{Stanford}}(2017)}]{Mezei2017}%
  \BibitemOpen
  \bibfield  {author} {\bibinfo {author} {\bibfnamefont {M.}~\bibnamefont
  {{Mezei}}}\ and\ \bibinfo {author} {\bibfnamefont {D.}~\bibnamefont
  {{Stanford}}},\ }\href {\doibase 10.1007/JHEP05(2017)065} {\bibfield
  {journal} {\bibinfo  {journal} {Journal of High Energy Physics}\ }\textbf
  {\bibinfo {volume} {5}},\ \bibinfo {eid} {65} (\bibinfo {year} {2017})},\
  \Eprint {http://arxiv.org/abs/1608.05101} {arXiv:1608.05101 [hep-th]}
  \BibitemShut {NoStop}%
\bibitem [{\citenamefont {{Nahum}}\ \emph {et~al.}(2017)\citenamefont
  {{Nahum}}, \citenamefont {{Ruhman}}, \citenamefont {{Vijay}},\ and\
  \citenamefont {{Haah}}}]{nahum2017KPZ}%
  \BibitemOpen
  \bibfield  {author} {\bibinfo {author} {\bibfnamefont {A.}~\bibnamefont
  {{Nahum}}}, \bibinfo {author} {\bibfnamefont {J.}~\bibnamefont {{Ruhman}}},
  \bibinfo {author} {\bibfnamefont {S.}~\bibnamefont {{Vijay}}}, \ and\
  \bibinfo {author} {\bibfnamefont {J.}~\bibnamefont {{Haah}}},\ }\href
  {\doibase 10.1103/PhysRevX.7.031016} {\bibfield  {journal} {\bibinfo
  {journal} {Physical Review X}\ }\textbf {\bibinfo {volume} {7}},\ \bibinfo
  {eid} {031016} (\bibinfo {year} {2017})},\ \Eprint
  {http://arxiv.org/abs/1608.06950} {arXiv:1608.06950 [cond-mat.stat-mech]}
  \BibitemShut {NoStop}%
\bibitem [{\citenamefont {{Nahum}}\ \emph {et~al.}(2018)\citenamefont
  {{Nahum}}, \citenamefont {{Vijay}},\ and\ \citenamefont
  {{Haah}}}]{nahum2018operator}%
  \BibitemOpen
  \bibfield  {author} {\bibinfo {author} {\bibfnamefont {A.}~\bibnamefont
  {{Nahum}}}, \bibinfo {author} {\bibfnamefont {S.}~\bibnamefont {{Vijay}}}, \
  and\ \bibinfo {author} {\bibfnamefont {J.}~\bibnamefont {{Haah}}},\ }\href
  {\doibase 10.1103/PhysRevX.8.021014} {\bibfield  {journal} {\bibinfo
  {journal} {Physical Review X}\ }\textbf {\bibinfo {volume} {8}},\ \bibinfo
  {eid} {021014} (\bibinfo {year} {2018})},\ \Eprint
  {http://arxiv.org/abs/1705.08975} {arXiv:1705.08975 [cond-mat.str-el]}
  \BibitemShut {NoStop}%
\bibitem [{\citenamefont {{von Keyserlingk}}\ \emph {et~al.}(2018)\citenamefont
  {{von Keyserlingk}}, \citenamefont {{Rakovszky}}, \citenamefont
  {{Pollmann}},\ and\ \citenamefont {{Sondhi}}}]{keyserlingk2018operator}%
  \BibitemOpen
  \bibfield  {author} {\bibinfo {author} {\bibfnamefont {C.~W.}\ \bibnamefont
  {{von Keyserlingk}}}, \bibinfo {author} {\bibfnamefont {T.}~\bibnamefont
  {{Rakovszky}}}, \bibinfo {author} {\bibfnamefont {F.}~\bibnamefont
  {{Pollmann}}}, \ and\ \bibinfo {author} {\bibfnamefont {S.~L.}\ \bibnamefont
  {{Sondhi}}},\ }\href {\doibase 10.1103/PhysRevX.8.021013} {\bibfield
  {journal} {\bibinfo  {journal} {Physical Review X}\ }\textbf {\bibinfo
  {volume} {8}},\ \bibinfo {eid} {021013} (\bibinfo {year} {2018})},\ \Eprint
  {http://arxiv.org/abs/1705.08910} {arXiv:1705.08910 [cond-mat.str-el]}
  \BibitemShut {NoStop}%
\bibitem [{\citenamefont {{Wiseman}}(1996)}]{Wiseman1996}%
  \BibitemOpen
  \bibfield  {author} {\bibinfo {author} {\bibfnamefont {H.~M.}\ \bibnamefont
  {{Wiseman}}},\ }\href {\doibase 10.1088/1355-5111/8/1/015} {\bibfield
  {journal} {\bibinfo  {journal} {Quantum and Semiclassical Optics, Volume 8,
  Issue 1, pp.~205-222 (1996).}\ }\textbf {\bibinfo {volume} {8}},\ \bibinfo
  {pages} {205} (\bibinfo {year} {1996})},\ \Eprint
  {http://arxiv.org/abs/quant-ph/0302080} {quant-ph/0302080} \BibitemShut
  {NoStop}%
\bibitem [{\citenamefont {Dhar}\ and\ \citenamefont
  {Dasgupta}(2016)}]{dasgupta2016measurement}%
  \BibitemOpen
  \bibfield  {author} {\bibinfo {author} {\bibfnamefont {S.}~\bibnamefont
  {Dhar}}\ and\ \bibinfo {author} {\bibfnamefont {S.}~\bibnamefont
  {Dasgupta}},\ }\href {\doibase 10.1103/PhysRevA.93.050103} {\  (\bibinfo
  {year} {2016}),\ 10.1103/PhysRevA.93.050103},\ \Eprint
  {http://arxiv.org/abs/1603.03561} {arXiv:1603.03561} \BibitemShut {NoStop}%
\bibitem [{\citenamefont {Elliott}\ \emph {et~al.}(2015)\citenamefont
  {Elliott}, \citenamefont {Kozlowski}, \citenamefont {Caballero-Benitez},\
  and\ \citenamefont {Mekhov}}]{elliott2015cold}%
  \BibitemOpen
  \bibfield  {author} {\bibinfo {author} {\bibfnamefont {T.~J.}\ \bibnamefont
  {Elliott}}, \bibinfo {author} {\bibfnamefont {W.}~\bibnamefont {Kozlowski}},
  \bibinfo {author} {\bibfnamefont {S.~F.}\ \bibnamefont {Caballero-Benitez}},
  \ and\ \bibinfo {author} {\bibfnamefont {I.~B.}\ \bibnamefont {Mekhov}},\
  }\href {\doibase 10.1103/PhysRevLett.114.113604} {\bibfield  {journal}
  {\bibinfo  {journal} {Physical Review Letters}\ }\textbf {\bibinfo {volume}
  {114}},\ \bibinfo {pages} {113604} (\bibinfo {year} {2015})},\ \Eprint
  {http://arxiv.org/abs/1412.4680} {arXiv:1412.4680} \BibitemShut {NoStop}%
\bibitem [{\citenamefont {Mazzucchi}\ \emph {et~al.}(2016)\citenamefont
  {Mazzucchi}, \citenamefont {Kozlowski}, \citenamefont {Caballero-Benitez},
  \citenamefont {Elliott},\ and\ \citenamefont {Mekhov}}]{elliott2016cold}%
  \BibitemOpen
  \bibfield  {author} {\bibinfo {author} {\bibfnamefont {G.}~\bibnamefont
  {Mazzucchi}}, \bibinfo {author} {\bibfnamefont {W.}~\bibnamefont
  {Kozlowski}}, \bibinfo {author} {\bibfnamefont {S.~F.}\ \bibnamefont
  {Caballero-Benitez}}, \bibinfo {author} {\bibfnamefont {T.~J.}\ \bibnamefont
  {Elliott}}, \ and\ \bibinfo {author} {\bibfnamefont {I.~B.}\ \bibnamefont
  {Mekhov}},\ }\href {\doibase 10.1103/PhysRevA.93.023632} {\bibfield
  {journal} {\bibinfo  {journal} {Physical Review A}\ }\textbf {\bibinfo
  {volume} {93}},\ \bibinfo {pages} {023632} (\bibinfo {year} {2016})},\
  \Eprint {http://arxiv.org/abs/1503.08710} {arXiv:1503.08710} \BibitemShut
  {NoStop}%
\bibitem [{\citenamefont {{Misra}}\ and\ \citenamefont
  {{Sudarshan}}(1977)}]{Misra1977zeno}%
  \BibitemOpen
  \bibfield  {author} {\bibinfo {author} {\bibfnamefont {B.}~\bibnamefont
  {{Misra}}}\ and\ \bibinfo {author} {\bibfnamefont {E.~C.~G.}\ \bibnamefont
  {{Sudarshan}}},\ }\href {\doibase 10.1063/1.523304} {\bibfield  {journal}
  {\bibinfo  {journal} {Journal of Mathematical Physics}\ }\textbf {\bibinfo
  {volume} {18}},\ \bibinfo {pages} {756} (\bibinfo {year} {1977})}\BibitemShut
  {NoStop}%
\bibitem [{\citenamefont {Breuer}\ \emph {et~al.}(2002)\citenamefont {Breuer},
  \citenamefont {Petruccione} \emph {et~al.}}]{Breuer2002theory}%
  \BibitemOpen
  \bibfield  {author} {\bibinfo {author} {\bibfnamefont {H.-P.}\ \bibnamefont
  {Breuer}}, \bibinfo {author} {\bibfnamefont {F.}~\bibnamefont {Petruccione}},
   \emph {et~al.},\ }\href@noop {} {\emph {\bibinfo {title} {The theory of open
  quantum systems}}}\ (\bibinfo  {publisher} {Oxford University Press on
  Demand},\ \bibinfo {year} {2002})\BibitemShut {NoStop}%
\bibitem [{\citenamefont {{Cao}}\ \emph {et~al.}(2018)\citenamefont {{Cao}},
  \citenamefont {{Tilloy}},\ and\ \citenamefont {{De
  Luca}}}]{cao2018monitoring}%
  \BibitemOpen
  \bibfield  {author} {\bibinfo {author} {\bibfnamefont {X.}~\bibnamefont
  {{Cao}}}, \bibinfo {author} {\bibfnamefont {A.}~\bibnamefont {{Tilloy}}}, \
  and\ \bibinfo {author} {\bibfnamefont {A.}~\bibnamefont {{De Luca}}},\
  }\href@noop {} {\bibfield  {journal} {\bibinfo  {journal} {ArXiv e-prints}\ }
  (\bibinfo {year} {2018})},\ \Eprint {http://arxiv.org/abs/1804.04638}
  {arXiv:1804.04638 [cond-mat.stat-mech]} \BibitemShut {NoStop}%
\bibitem [{\citenamefont {{Grover}}\ and\ \citenamefont
  {{Fisher}}(2015)}]{Grover2015}%
  \BibitemOpen
  \bibfield  {author} {\bibinfo {author} {\bibfnamefont {T.}~\bibnamefont
  {{Grover}}}\ and\ \bibinfo {author} {\bibfnamefont {M.~P.~A.}\ \bibnamefont
  {{Fisher}}},\ }\href {\doibase 10.1103/PhysRevA.92.042308} {\bibfield
  {journal} {\bibinfo  {journal} {\pra}\ }\textbf {\bibinfo {volume} {92}},\
  \bibinfo {eid} {042308} (\bibinfo {year} {2015})},\ \Eprint
  {http://arxiv.org/abs/1412.3534} {arXiv:1412.3534 [cond-mat.stat-mech]}
  \BibitemShut {NoStop}%
\bibitem [{\citenamefont {Fisher}\ and\ \citenamefont
  {Radzihovsky}(2018)}]{mpaf1707qds}%
  \BibitemOpen
  \bibfield  {author} {\bibinfo {author} {\bibfnamefont {M.~P.~A.}\
  \bibnamefont {Fisher}}\ and\ \bibinfo {author} {\bibfnamefont
  {L.}~\bibnamefont {Radzihovsky}},\ }\href {\doibase 10.1073/pnas.1718402115}
  {\bibfield  {journal} {\bibinfo  {journal} {Proceedings of the National
  Academy of Sciences}\ }\textbf {\bibinfo {volume} {115}},\ \bibinfo {pages}
  {E4551} (\bibinfo {year} {2018})},\ \Eprint {http://arxiv.org/abs/1707.05320}
  {arXiv:1707.05320 [physics.chem-ph]} \BibitemShut {NoStop}%
\bibitem [{\citenamefont {{Fisher}}(2015)}]{mpaf1508qcog}%
  \BibitemOpen
  \bibfield  {author} {\bibinfo {author} {\bibfnamefont {M.~P.~A.}\
  \bibnamefont {{Fisher}}},\ }\href {\doibase 10.1016/j.aop.2015.08.020}
  {\bibfield  {journal} {\bibinfo  {journal} {Annals of Physics}\ }\textbf
  {\bibinfo {volume} {362}},\ \bibinfo {pages} {593} (\bibinfo {year}
  {2015})},\ \Eprint {http://arxiv.org/abs/1508.05929} {arXiv:1508.05929
  [q-bio.NC]} \BibitemShut {NoStop}%
\bibitem [{\citenamefont {{Swift}}\ \emph {et~al.}(2018)\citenamefont
  {{Swift}}, \citenamefont {{Van de Walle}},\ and\ \citenamefont
  {{Fisher}}}]{swift2018posner}%
  \BibitemOpen
  \bibfield  {author} {\bibinfo {author} {\bibfnamefont {M.~W.}\ \bibnamefont
  {{Swift}}}, \bibinfo {author} {\bibfnamefont {C.~G.}\ \bibnamefont {{Van de
  Walle}}}, \ and\ \bibinfo {author} {\bibfnamefont {M.~P.~A.}\ \bibnamefont
  {{Fisher}}},\ }\href {\doibase 10.1039/C7CP07720C} {\bibfield  {journal}
  {\bibinfo  {journal} {Physical Chemistry Chemical Physics (Incorporating
  Faraday Transactions)}\ }\textbf {\bibinfo {volume} {20}},\ \bibinfo {pages}
  {12373} (\bibinfo {year} {2018})},\ \Eprint {http://arxiv.org/abs/1711.05899}
  {arXiv:1711.05899 [physics.chem-ph]} \BibitemShut {NoStop}%
\bibitem [{\citenamefont {Li}\ and\ \citenamefont
  {Fisher}()}]{YaodongMPAF2018Cat}%
  \BibitemOpen
  \bibfield  {author} {\bibinfo {author} {\bibfnamefont {Y.}~\bibnamefont
  {Li}}\ and\ \bibinfo {author} {\bibfnamefont {M.~P.~A.}\ \bibnamefont
  {Fisher}},\ }\href@noop {} {\ }\Eprint {http://arxiv.org/abs/in preparation}
  {in preparation} \BibitemShut {NoStop}%
\bibitem [{\citenamefont {{Yunger Halpern}}\ and\ \citenamefont
  {{Crosson}}(2017)}]{halpern1711qiqcog}%
  \BibitemOpen
  \bibfield  {author} {\bibinfo {author} {\bibfnamefont {N.}~\bibnamefont
  {{Yunger Halpern}}}\ and\ \bibinfo {author} {\bibfnamefont {E.}~\bibnamefont
  {{Crosson}}},\ }\href@noop {} {\bibfield  {journal} {\bibinfo  {journal}
  {ArXiv e-prints}\ } (\bibinfo {year} {2017})},\ \Eprint
  {http://arxiv.org/abs/1711.04801} {arXiv:1711.04801 [quant-ph]} \BibitemShut
  {NoStop}%
\bibitem [{\citenamefont {{Nielsen}}\ and\ \citenamefont
  {{Chuang}}(2010)}]{nielsen2010qiqc}%
  \BibitemOpen
  \bibfield  {author} {\bibinfo {author} {\bibfnamefont {M.~A.}\ \bibnamefont
  {{Nielsen}}}\ and\ \bibinfo {author} {\bibfnamefont {I.~L.}\ \bibnamefont
  {{Chuang}}},\ }\href@noop {} {\emph {\bibinfo {title} {Quantum Computation
  and Quantum Information, by Michael A.~Nielsen , Isaac L.~Chuang, Cambridge,
  UK: Cambridge University Press, 2010}}}\ (\bibinfo {year} {2010})\BibitemShut
  {NoStop}%
\bibitem [{\citenamefont {Mehta}(2004)}]{mehta2004matrices}%
  \BibitemOpen
  \bibfield  {author} {\bibinfo {author} {\bibfnamefont {M.~L.}\ \bibnamefont
  {Mehta}},\ }\href@noop {} {\emph {\bibinfo {title} {Random Matrices}}},\
  \bibinfo {edition} {3rd}\ ed.\ (\bibinfo {year} {2004})\BibitemShut {NoStop}%
\bibitem [{\citenamefont {Forrester}(2010)}]{loggasrandommatrices}%
  \BibitemOpen
  \bibfield  {author} {\bibinfo {author} {\bibfnamefont {P.}~\bibnamefont
  {Forrester}},\ }\href {http://www.jstor.org/stable/j.ctt7t5vq} {\emph
  {\bibinfo {title} {Log-Gases and Random Matrices (LMS-34)}}}\ (\bibinfo
  {publisher} {Princeton University Press},\ \bibinfo {year}
  {2010})\BibitemShut {NoStop}%
\bibitem [{\citenamefont {{Gottesman}}(1998)}]{gottesman9807heisenberg}%
  \BibitemOpen
  \bibfield  {author} {\bibinfo {author} {\bibfnamefont {D.}~\bibnamefont
  {{Gottesman}}},\ }\href@noop {} {\bibfield  {journal} {\bibinfo  {journal}
  {eprint arXiv:quant-ph/9807006}\ } (\bibinfo {year} {1998})},\ \Eprint
  {http://arxiv.org/abs/quant-ph/9807006} {quant-ph/9807006} \BibitemShut
  {NoStop}%
\bibitem [{\citenamefont {{Aaronson}}\ and\ \citenamefont
  {{Gottesman}}(2004)}]{aaronson0406chp}%
  \BibitemOpen
  \bibfield  {author} {\bibinfo {author} {\bibfnamefont {S.}~\bibnamefont
  {{Aaronson}}}\ and\ \bibinfo {author} {\bibfnamefont {D.}~\bibnamefont
  {{Gottesman}}},\ }\href {\doibase 10.1103/PhysRevA.70.052328} {\bibfield
  {journal} {\bibinfo  {journal} {\pra}\ }\textbf {\bibinfo {volume} {70}},\
  \bibinfo {eid} {052328} (\bibinfo {year} {2004})},\ \Eprint
  {http://arxiv.org/abs/quant-ph/0406196} {quant-ph/0406196} \BibitemShut
  {NoStop}%
\bibitem [{\citenamefont {DiVincenzo}\ \emph {et~al.}(2002)\citenamefont
  {DiVincenzo}, \citenamefont {Leung},\ and\ \citenamefont
  {Terhal}}]{DiVincenzo2002hiding}%
  \BibitemOpen
  \bibfield  {author} {\bibinfo {author} {\bibfnamefont {D.~P.}\ \bibnamefont
  {DiVincenzo}}, \bibinfo {author} {\bibfnamefont {D.~W.}\ \bibnamefont
  {Leung}}, \ and\ \bibinfo {author} {\bibfnamefont {B.~M.}\ \bibnamefont
  {Terhal}},\ }\href {\doibase 10.1109/18.985948} {\bibfield  {journal}
  {\bibinfo  {journal} {IEEE Transactions on Information Theory}\ }\textbf
  {\bibinfo {volume} {48}},\ \bibinfo {pages} {580} (\bibinfo {year} {2002})},\
  \Eprint {http://arxiv.org/abs/quant-ph/0103098} {quant-ph/0103098}
  \BibitemShut {NoStop}%
\bibitem [{\citenamefont {{Hamma}}\ \emph
  {et~al.}(2005{\natexlab{a}})\citenamefont {{Hamma}}, \citenamefont
  {{Ionicioiu}},\ and\ \citenamefont {{Zanardi}}}]{hamma2005bipartite}%
  \BibitemOpen
  \bibfield  {author} {\bibinfo {author} {\bibfnamefont {A.}~\bibnamefont
  {{Hamma}}}, \bibinfo {author} {\bibfnamefont {R.}~\bibnamefont
  {{Ionicioiu}}}, \ and\ \bibinfo {author} {\bibfnamefont {P.}~\bibnamefont
  {{Zanardi}}},\ }\href {\doibase 10.1103/PhysRevA.71.022315} {\bibfield
  {journal} {\bibinfo  {journal} {\pra}\ }\textbf {\bibinfo {volume} {71}},\
  \bibinfo {eid} {022315} (\bibinfo {year} {2005}{\natexlab{a}})},\ \Eprint
  {http://arxiv.org/abs/quant-ph/0409073} {quant-ph/0409073} \BibitemShut
  {NoStop}%
\bibitem [{\citenamefont {{Hamma}}\ \emph
  {et~al.}(2005{\natexlab{b}})\citenamefont {{Hamma}}, \citenamefont
  {{Ionicioiu}},\ and\ \citenamefont {{Zanardi}}}]{hamma2005entanglmeent}%
  \BibitemOpen
  \bibfield  {author} {\bibinfo {author} {\bibfnamefont {A.}~\bibnamefont
  {{Hamma}}}, \bibinfo {author} {\bibfnamefont {R.}~\bibnamefont
  {{Ionicioiu}}}, \ and\ \bibinfo {author} {\bibfnamefont {P.}~\bibnamefont
  {{Zanardi}}},\ }\href {\doibase 10.1016/j.physleta.2005.01.060} {\bibfield
  {journal} {\bibinfo  {journal} {Physics Letters A}\ }\textbf {\bibinfo
  {volume} {337}},\ \bibinfo {pages} {22} (\bibinfo {year}
  {2005}{\natexlab{b}})},\ \Eprint {http://arxiv.org/abs/quant-ph/0406202}
  {quant-ph/0406202} \BibitemShut {NoStop}%
\bibitem [{Note1()}]{Note1}%
  \BibitemOpen
  \bibinfo {note} {We note that in Refs. \cite {Vasseur2018, nahum2018hybrid},
  a different scaling form for entanglement entropy, $S_A(p, L_A) = a(p) \log
  L_A + b(p) L_A$ was used for data collapse.}\BibitemShut {Stop}%
\bibitem [{\citenamefont {{Chen}}\ \emph {et~al.}(2015)\citenamefont {{Chen}},
  \citenamefont {{Yu}}, \citenamefont {{Cho}}, \citenamefont {{Clark}},\ and\
  \citenamefont {{Fradkin}}}]{Chen2015}%
  \BibitemOpen
  \bibfield  {author} {\bibinfo {author} {\bibfnamefont {X.}~\bibnamefont
  {{Chen}}}, \bibinfo {author} {\bibfnamefont {X.}~\bibnamefont {{Yu}}},
  \bibinfo {author} {\bibfnamefont {G.~Y.}\ \bibnamefont {{Cho}}}, \bibinfo
  {author} {\bibfnamefont {B.~K.}\ \bibnamefont {{Clark}}}, \ and\ \bibinfo
  {author} {\bibfnamefont {E.}~\bibnamefont {{Fradkin}}},\ }\href {\doibase
  10.1103/PhysRevB.92.214204} {\bibfield  {journal} {\bibinfo  {journal}
  {\prb}\ }\textbf {\bibinfo {volume} {92}},\ \bibinfo {eid} {214204} (\bibinfo
  {year} {2015})},\ \Eprint {http://arxiv.org/abs/1509.03890} {arXiv:1509.03890
  [cond-mat.dis-nn]} \BibitemShut {NoStop}%
\bibitem [{\citenamefont {{Vasseur}}\ \emph {et~al.}(2018)\citenamefont
  {{Vasseur}}, \citenamefont {{Potter}}, \citenamefont {{You}},\ and\
  \citenamefont {{Ludwig}}}]{Vasseur2018}%
  \BibitemOpen
  \bibfield  {author} {\bibinfo {author} {\bibfnamefont {R.}~\bibnamefont
  {{Vasseur}}}, \bibinfo {author} {\bibfnamefont {A.~C.}\ \bibnamefont
  {{Potter}}}, \bibinfo {author} {\bibfnamefont {Y.-Z.}\ \bibnamefont {{You}}},
  \ and\ \bibinfo {author} {\bibfnamefont {A.~W.~W.}\ \bibnamefont
  {{Ludwig}}},\ }\href@noop {} {\bibfield  {journal} {\bibinfo  {journal}
  {ArXiv e-prints}\ } (\bibinfo {year} {2018})},\ \Eprint
  {http://arxiv.org/abs/1807.07082} {arXiv:1807.07082 [cond-mat.stat-mech]}
  \BibitemShut {NoStop}%
\bibitem [{\citenamefont {{Kj{\"a}ll}}\ \emph {et~al.}(2014)\citenamefont
  {{Kj{\"a}ll}}, \citenamefont {{Bardarson}},\ and\ \citenamefont
  {{Pollmann}}}]{Kjall2014}%
  \BibitemOpen
  \bibfield  {author} {\bibinfo {author} {\bibfnamefont {J.~A.}\ \bibnamefont
  {{Kj{\"a}ll}}}, \bibinfo {author} {\bibfnamefont {J.~H.}\ \bibnamefont
  {{Bardarson}}}, \ and\ \bibinfo {author} {\bibfnamefont {F.}~\bibnamefont
  {{Pollmann}}},\ }\href {\doibase 10.1103/PhysRevLett.113.107204} {\bibfield
  {journal} {\bibinfo  {journal} {Physical Review Letters}\ }\textbf {\bibinfo
  {volume} {113}},\ \bibinfo {eid} {107204} (\bibinfo {year} {2014})},\ \Eprint
  {http://arxiv.org/abs/1403.1568} {arXiv:1403.1568 [cond-mat.str-el]}
  \BibitemShut {NoStop}%
\bibitem [{\citenamefont {{Luitz}}\ \emph {et~al.}(2015)\citenamefont
  {{Luitz}}, \citenamefont {{Laflorencie}},\ and\ \citenamefont
  {{Alet}}}]{Luitz2015}%
  \BibitemOpen
  \bibfield  {author} {\bibinfo {author} {\bibfnamefont {D.~J.}\ \bibnamefont
  {{Luitz}}}, \bibinfo {author} {\bibfnamefont {N.}~\bibnamefont
  {{Laflorencie}}}, \ and\ \bibinfo {author} {\bibfnamefont {F.}~\bibnamefont
  {{Alet}}},\ }\href {\doibase 10.1103/PhysRevB.91.081103} {\bibfield
  {journal} {\bibinfo  {journal} {\prb}\ }\textbf {\bibinfo {volume} {91}},\
  \bibinfo {eid} {081103} (\bibinfo {year} {2015})},\ \Eprint
  {http://arxiv.org/abs/1411.0660} {arXiv:1411.0660 [cond-mat.dis-nn]}
  \BibitemShut {NoStop}%
\bibitem [{\citenamefont {{Vosk}}\ \emph {et~al.}(2015)\citenamefont {{Vosk}},
  \citenamefont {{Huse}},\ and\ \citenamefont {{Altman}}}]{Vosk2015}%
  \BibitemOpen
  \bibfield  {author} {\bibinfo {author} {\bibfnamefont {R.}~\bibnamefont
  {{Vosk}}}, \bibinfo {author} {\bibfnamefont {D.~A.}\ \bibnamefont {{Huse}}},
  \ and\ \bibinfo {author} {\bibfnamefont {E.}~\bibnamefont {{Altman}}},\
  }\href {\doibase 10.1103/PhysRevX.5.031032} {\bibfield  {journal} {\bibinfo
  {journal} {Physical Review X}\ }\textbf {\bibinfo {volume} {5}},\ \bibinfo
  {eid} {031032} (\bibinfo {year} {2015})},\ \Eprint
  {http://arxiv.org/abs/1412.3117} {arXiv:1412.3117 [cond-mat.dis-nn]}
  \BibitemShut {NoStop}%
\bibitem [{\citenamefont {{Potter}}\ \emph {et~al.}(2015)\citenamefont
  {{Potter}}, \citenamefont {{Vasseur}},\ and\ \citenamefont
  {{Parameswaran}}}]{Potter2015}%
  \BibitemOpen
  \bibfield  {author} {\bibinfo {author} {\bibfnamefont {A.~C.}\ \bibnamefont
  {{Potter}}}, \bibinfo {author} {\bibfnamefont {R.}~\bibnamefont {{Vasseur}}},
  \ and\ \bibinfo {author} {\bibfnamefont {S.~A.}\ \bibnamefont
  {{Parameswaran}}},\ }\href {\doibase 10.1103/PhysRevX.5.031033} {\bibfield
  {journal} {\bibinfo  {journal} {Physical Review X}\ }\textbf {\bibinfo
  {volume} {5}},\ \bibinfo {eid} {031033} (\bibinfo {year} {2015})},\ \Eprint
  {http://arxiv.org/abs/1501.03501} {arXiv:1501.03501 [cond-mat.dis-nn]}
  \BibitemShut {NoStop}%
\bibitem [{\citenamefont {{Serbyn}}\ \emph {et~al.}(2015)\citenamefont
  {{Serbyn}}, \citenamefont {{Papi{\'c}}},\ and\ \citenamefont
  {{Abanin}}}]{Serbyn2015}%
  \BibitemOpen
  \bibfield  {author} {\bibinfo {author} {\bibfnamefont {M.}~\bibnamefont
  {{Serbyn}}}, \bibinfo {author} {\bibfnamefont {Z.}~\bibnamefont
  {{Papi{\'c}}}}, \ and\ \bibinfo {author} {\bibfnamefont {D.~A.}\ \bibnamefont
  {{Abanin}}},\ }\href {\doibase 10.1103/PhysRevX.5.041047} {\bibfield
  {journal} {\bibinfo  {journal} {Physical Review X}\ }\textbf {\bibinfo
  {volume} {5}},\ \bibinfo {eid} {041047} (\bibinfo {year} {2015})},\ \Eprint
  {http://arxiv.org/abs/1507.01635} {arXiv:1507.01635 [cond-mat.dis-nn]}
  \BibitemShut {NoStop}%
\bibitem [{\citenamefont {{Zhang}}\ \emph {et~al.}(2016)\citenamefont
  {{Zhang}}, \citenamefont {{Zhao}}, \citenamefont {{Devakul}},\ and\
  \citenamefont {{Huse}}}]{Zhang2016}%
  \BibitemOpen
  \bibfield  {author} {\bibinfo {author} {\bibfnamefont {L.}~\bibnamefont
  {{Zhang}}}, \bibinfo {author} {\bibfnamefont {B.}~\bibnamefont {{Zhao}}},
  \bibinfo {author} {\bibfnamefont {T.}~\bibnamefont {{Devakul}}}, \ and\
  \bibinfo {author} {\bibfnamefont {D.~A.}\ \bibnamefont {{Huse}}},\ }\href
  {\doibase 10.1103/PhysRevB.93.224201} {\bibfield  {journal} {\bibinfo
  {journal} {\prb}\ }\textbf {\bibinfo {volume} {93}},\ \bibinfo {eid} {224201}
  (\bibinfo {year} {2016})},\ \Eprint {http://arxiv.org/abs/1603.02296}
  {arXiv:1603.02296 [cond-mat.stat-mech]} \BibitemShut {NoStop}%
\bibitem [{\citenamefont {{Dumitrescu}}\ \emph {et~al.}(2017)\citenamefont
  {{Dumitrescu}}, \citenamefont {{Vasseur}},\ and\ \citenamefont
  {{Potter}}}]{Dumitrescu2017}%
  \BibitemOpen
  \bibfield  {author} {\bibinfo {author} {\bibfnamefont {P.~T.}\ \bibnamefont
  {{Dumitrescu}}}, \bibinfo {author} {\bibfnamefont {R.}~\bibnamefont
  {{Vasseur}}}, \ and\ \bibinfo {author} {\bibfnamefont {A.~C.}\ \bibnamefont
  {{Potter}}},\ }\href {\doibase 10.1103/PhysRevLett.119.110604} {\bibfield
  {journal} {\bibinfo  {journal} {Physical Review Letters}\ }\textbf {\bibinfo
  {volume} {119}},\ \bibinfo {eid} {110604} (\bibinfo {year} {2017})},\ \Eprint
  {http://arxiv.org/abs/1701.04827} {arXiv:1701.04827 [cond-mat.dis-nn]}
  \BibitemShut {NoStop}%
\bibitem [{\citenamefont {{Khemani}}\ \emph {et~al.}(2017)\citenamefont
  {{Khemani}}, \citenamefont {{Lim}}, \citenamefont {{Sheng}},\ and\
  \citenamefont {{Huse}}}]{Khemani2017}%
  \BibitemOpen
  \bibfield  {author} {\bibinfo {author} {\bibfnamefont {V.}~\bibnamefont
  {{Khemani}}}, \bibinfo {author} {\bibfnamefont {S.~P.}\ \bibnamefont
  {{Lim}}}, \bibinfo {author} {\bibfnamefont {D.~N.}\ \bibnamefont {{Sheng}}},
  \ and\ \bibinfo {author} {\bibfnamefont {D.~A.}\ \bibnamefont {{Huse}}},\
  }\href {\doibase 10.1103/PhysRevX.7.021013} {\bibfield  {journal} {\bibinfo
  {journal} {Physical Review X}\ }\textbf {\bibinfo {volume} {7}},\ \bibinfo
  {eid} {021013} (\bibinfo {year} {2017})},\ \Eprint
  {http://arxiv.org/abs/1607.05756} {arXiv:1607.05756 [cond-mat.dis-nn]}
  \BibitemShut {NoStop}%
\bibitem [{\citenamefont {{Pal}}\ and\ \citenamefont {{Huse}}(2010)}]{Pal2010}%
  \BibitemOpen
  \bibfield  {author} {\bibinfo {author} {\bibfnamefont {A.}~\bibnamefont
  {{Pal}}}\ and\ \bibinfo {author} {\bibfnamefont {D.~A.}\ \bibnamefont
  {{Huse}}},\ }\href {\doibase 10.1103/PhysRevB.82.174411} {\bibfield
  {journal} {\bibinfo  {journal} {\prb}\ }\textbf {\bibinfo {volume} {82}},\
  \bibinfo {eid} {174411} (\bibinfo {year} {2010})},\ \Eprint
  {http://arxiv.org/abs/1010.1992} {arXiv:1010.1992 [cond-mat.dis-nn]}
  \BibitemShut {NoStop}%
\bibitem [{\citenamefont {{Bauer}}\ and\ \citenamefont
  {{Nayak}}(2013)}]{Bauer2013}%
  \BibitemOpen
  \bibfield  {author} {\bibinfo {author} {\bibfnamefont {B.}~\bibnamefont
  {{Bauer}}}\ and\ \bibinfo {author} {\bibfnamefont {C.}~\bibnamefont
  {{Nayak}}},\ }\href {\doibase 10.1088/1742-5468/2013/09/P09005} {\bibfield
  {journal} {\bibinfo  {journal} {Journal of Statistical Mechanics: Theory and
  Experiment}\ }\textbf {\bibinfo {volume} {9}},\ \bibinfo {eid} {09005}
  (\bibinfo {year} {2013})},\ \Eprint {http://arxiv.org/abs/1306.5753}
  {arXiv:1306.5753 [cond-mat.dis-nn]} \BibitemShut {NoStop}%
\bibitem [{\citenamefont {{Nandkishore}}\ and\ \citenamefont
  {{Huse}}(2015)}]{Nandkishore2015}%
  \BibitemOpen
  \bibfield  {author} {\bibinfo {author} {\bibfnamefont {R.}~\bibnamefont
  {{Nandkishore}}}\ and\ \bibinfo {author} {\bibfnamefont {D.~A.}\ \bibnamefont
  {{Huse}}},\ }\href {\doibase 10.1146/annurev-conmatphys-031214-014726}
  {\bibfield  {journal} {\bibinfo  {journal} {Annual Review of Condensed Matter
  Physics}\ }\textbf {\bibinfo {volume} {6}},\ \bibinfo {pages} {15} (\bibinfo
  {year} {2015})},\ \Eprint {http://arxiv.org/abs/1404.0686} {arXiv:1404.0686
  [cond-mat.stat-mech]} \BibitemShut {NoStop}%
\bibitem [{\citenamefont {{Deutsch}}(1991)}]{Deutsch1991}%
  \BibitemOpen
  \bibfield  {author} {\bibinfo {author} {\bibfnamefont {J.~M.}\ \bibnamefont
  {{Deutsch}}},\ }\href {\doibase 10.1103/PhysRevA.43.2046} {\bibfield
  {journal} {\bibinfo  {journal} {\pra}\ }\textbf {\bibinfo {volume} {43}},\
  \bibinfo {pages} {2046} (\bibinfo {year} {1991})}\BibitemShut {NoStop}%
\bibitem [{\citenamefont {{Srednicki}}(1994)}]{Srednicki1994}%
  \BibitemOpen
  \bibfield  {author} {\bibinfo {author} {\bibfnamefont {M.}~\bibnamefont
  {{Srednicki}}},\ }\href {\doibase 10.1103/PhysRevE.50.888} {\bibfield
  {journal} {\bibinfo  {journal} {\pre}\ }\textbf {\bibinfo {volume} {50}},\
  \bibinfo {pages} {888} (\bibinfo {year} {1994})},\ \Eprint
  {http://arxiv.org/abs/cond-mat/9403051} {cond-mat/9403051} \BibitemShut
  {NoStop}%
\bibitem [{\citenamefont {{Griffiths}}(1969)}]{Griffiths1969}%
  \BibitemOpen
  \bibfield  {author} {\bibinfo {author} {\bibfnamefont {R.~B.}\ \bibnamefont
  {{Griffiths}}},\ }\href {\doibase 10.1103/PhysRevLett.23.17} {\bibfield
  {journal} {\bibinfo  {journal} {Physical Review Letters}\ }\textbf {\bibinfo
  {volume} {23}},\ \bibinfo {pages} {17} (\bibinfo {year} {1969})}\BibitemShut
  {NoStop}%
\bibitem [{\citenamefont {{Agarwal}}\ \emph {et~al.}(2015)\citenamefont
  {{Agarwal}}, \citenamefont {{Gopalakrishnan}}, \citenamefont {{Knap}},
  \citenamefont {{M{\"u}ller}},\ and\ \citenamefont {{Demler}}}]{Agarwal2015}%
  \BibitemOpen
  \bibfield  {author} {\bibinfo {author} {\bibfnamefont {K.}~\bibnamefont
  {{Agarwal}}}, \bibinfo {author} {\bibfnamefont {S.}~\bibnamefont
  {{Gopalakrishnan}}}, \bibinfo {author} {\bibfnamefont {M.}~\bibnamefont
  {{Knap}}}, \bibinfo {author} {\bibfnamefont {M.}~\bibnamefont
  {{M{\"u}ller}}}, \ and\ \bibinfo {author} {\bibfnamefont {E.}~\bibnamefont
  {{Demler}}},\ }\href {\doibase 10.1103/PhysRevLett.114.160401} {\bibfield
  {journal} {\bibinfo  {journal} {Physical Review Letters}\ }\textbf {\bibinfo
  {volume} {114}},\ \bibinfo {eid} {160401} (\bibinfo {year} {2015})},\ \Eprint
  {http://arxiv.org/abs/1408.3413} {arXiv:1408.3413 [cond-mat.dis-nn]}
  \BibitemShut {NoStop}%
\bibitem [{\citenamefont {{Torres-Herrera}}\ and\ \citenamefont
  {{Santos}}(2015)}]{Torres2015}%
  \BibitemOpen
  \bibfield  {author} {\bibinfo {author} {\bibfnamefont {E.~J.}\ \bibnamefont
  {{Torres-Herrera}}}\ and\ \bibinfo {author} {\bibfnamefont {L.~F.}\
  \bibnamefont {{Santos}}},\ }\href {\doibase 10.1103/PhysRevB.92.014208}
  {\bibfield  {journal} {\bibinfo  {journal} {\prb}\ }\textbf {\bibinfo
  {volume} {92}},\ \bibinfo {eid} {014208} (\bibinfo {year} {2015})},\ \Eprint
  {http://arxiv.org/abs/1501.05662} {arXiv:1501.05662 [cond-mat.dis-nn]}
  \BibitemShut {NoStop}%
\bibitem [{\citenamefont {{Luitz}}\ \emph {et~al.}(2016)\citenamefont
  {{Luitz}}, \citenamefont {{Laflorencie}},\ and\ \citenamefont
  {{Alet}}}]{Luitz2016}%
  \BibitemOpen
  \bibfield  {author} {\bibinfo {author} {\bibfnamefont {D.~J.}\ \bibnamefont
  {{Luitz}}}, \bibinfo {author} {\bibfnamefont {N.}~\bibnamefont
  {{Laflorencie}}}, \ and\ \bibinfo {author} {\bibfnamefont {F.}~\bibnamefont
  {{Alet}}},\ }\href {\doibase 10.1103/PhysRevB.93.060201} {\bibfield
  {journal} {\bibinfo  {journal} {\prb}\ }\textbf {\bibinfo {volume} {93}},\
  \bibinfo {eid} {060201} (\bibinfo {year} {2016})},\ \Eprint
  {http://arxiv.org/abs/1511.05141} {arXiv:1511.05141 [cond-mat.dis-nn]}
  \BibitemShut {NoStop}%
\bibitem [{\citenamefont {{{\v Z}nidari{\v c}}}\ \emph
  {et~al.}(2008)\citenamefont {{{\v Z}nidari{\v c}}}, \citenamefont
  {{Prosen}},\ and\ \citenamefont {{Prelov{\v s}ek}}}]{Prosen2008}%
  \BibitemOpen
  \bibfield  {author} {\bibinfo {author} {\bibfnamefont {M.}~\bibnamefont {{{\v
  Z}nidari{\v c}}}}, \bibinfo {author} {\bibfnamefont {T.}~\bibnamefont
  {{Prosen}}}, \ and\ \bibinfo {author} {\bibfnamefont {P.}~\bibnamefont
  {{Prelov{\v s}ek}}},\ }\href {\doibase 10.1103/PhysRevB.77.064426} {\bibfield
   {journal} {\bibinfo  {journal} {\prb}\ }\textbf {\bibinfo {volume} {77}},\
  \bibinfo {eid} {064426} (\bibinfo {year} {2008})},\ \Eprint
  {http://arxiv.org/abs/0706.2539} {arXiv:0706.2539 [quant-ph]} \BibitemShut
  {NoStop}%
\bibitem [{\citenamefont {{Bardarson}}\ \emph {et~al.}(2012)\citenamefont
  {{Bardarson}}, \citenamefont {{Pollmann}},\ and\ \citenamefont
  {{Moore}}}]{Bardarson2012}%
  \BibitemOpen
  \bibfield  {author} {\bibinfo {author} {\bibfnamefont {J.~H.}\ \bibnamefont
  {{Bardarson}}}, \bibinfo {author} {\bibfnamefont {F.}~\bibnamefont
  {{Pollmann}}}, \ and\ \bibinfo {author} {\bibfnamefont {J.~E.}\ \bibnamefont
  {{Moore}}},\ }\href {\doibase 10.1103/PhysRevLett.109.017202} {\bibfield
  {journal} {\bibinfo  {journal} {Physical Review Letters}\ }\textbf {\bibinfo
  {volume} {109}},\ \bibinfo {eid} {017202} (\bibinfo {year} {2012})},\ \Eprint
  {http://arxiv.org/abs/1202.5532} {arXiv:1202.5532 [cond-mat.str-el]}
  \BibitemShut {NoStop}%
\bibitem [{\citenamefont {{Serbyn}}\ \emph {et~al.}(2013)\citenamefont
  {{Serbyn}}, \citenamefont {{Papi{\'c}}},\ and\ \citenamefont
  {{Abanin}}}]{Serbyn2013}%
  \BibitemOpen
  \bibfield  {author} {\bibinfo {author} {\bibfnamefont {M.}~\bibnamefont
  {{Serbyn}}}, \bibinfo {author} {\bibfnamefont {Z.}~\bibnamefont
  {{Papi{\'c}}}}, \ and\ \bibinfo {author} {\bibfnamefont {D.~A.}\ \bibnamefont
  {{Abanin}}},\ }\href {\doibase 10.1103/PhysRevLett.110.260601} {\bibfield
  {journal} {\bibinfo  {journal} {Physical Review Letters}\ }\textbf {\bibinfo
  {volume} {110}},\ \bibinfo {eid} {260601} (\bibinfo {year} {2013})},\ \Eprint
  {http://arxiv.org/abs/1304.4605} {arXiv:1304.4605 [cond-mat.str-el]}
  \BibitemShut {NoStop}%
\bibitem [{\citenamefont {{Grover}}(2014)}]{Grover2014}%
  \BibitemOpen
  \bibfield  {author} {\bibinfo {author} {\bibfnamefont {T.}~\bibnamefont
  {{Grover}}},\ }\href@noop {} {\bibfield  {journal} {\bibinfo  {journal}
  {ArXiv e-prints}\ } (\bibinfo {year} {2014})},\ \Eprint
  {http://arxiv.org/abs/1405.1471} {arXiv:1405.1471 [cond-mat.dis-nn]}
  \BibitemShut {NoStop}%
\bibitem [{\citenamefont {Chan}\ \emph {et~al.}(2018)\citenamefont {Chan},
  \citenamefont {Nandkishore}, \citenamefont {Pretko},\ and\ \citenamefont
  {Smith}}]{nandkishore2018hybrid}%
  \BibitemOpen
  \bibfield  {author} {\bibinfo {author} {\bibfnamefont {A.}~\bibnamefont
  {Chan}}, \bibinfo {author} {\bibfnamefont {R.~M.}\ \bibnamefont
  {Nandkishore}}, \bibinfo {author} {\bibfnamefont {M.}~\bibnamefont {Pretko}},
  \ and\ \bibinfo {author} {\bibfnamefont {G.}~\bibnamefont {Smith}},\ }\href
  {http://arxiv.org/abs/1808.05949} {\  (\bibinfo {year} {2018})},\ \Eprint
  {http://arxiv.org/abs/1808.05949} {arXiv:1808.05949} \BibitemShut {NoStop}%
\bibitem [{\citenamefont {Skinner}\ \emph {et~al.}(2018)\citenamefont
  {Skinner}, \citenamefont {Ruhman},\ and\ \citenamefont
  {Nahum}}]{nahum2018hybrid}%
  \BibitemOpen
  \bibfield  {author} {\bibinfo {author} {\bibfnamefont {B.}~\bibnamefont
  {Skinner}}, \bibinfo {author} {\bibfnamefont {J.}~\bibnamefont {Ruhman}}, \
  and\ \bibinfo {author} {\bibfnamefont {A.}~\bibnamefont {Nahum}},\ }\href
  {http://arxiv.org/abs/1808.05953} {\  (\bibinfo {year} {2018})},\ \Eprint
  {http://arxiv.org/abs/1808.05953} {arXiv:1808.05953} \BibitemShut {NoStop}%
\end{thebibliography}%


\end{document}